\RequirePackage{ifpdf}
\ifpdf % We are running pdfTeX in pdf mode
\documentclass[pdftex]{sigma}
\else
\documentclass{sigma}
\fi

\numberwithin{equation}{section}

\begin{document}

\allowdisplaybreaks

\renewcommand{\PaperNumber}{011}

\FirstPageHeading

\renewcommand{\thefootnote}{$\star$}

\ShortArticleName{Finite-Temperature Form Factors: a Review}

\ArticleName{Finite-Temperature Form Factors: a
Review\footnote{This paper is a contribution to the Proceedings of
the O'Raifeartaigh Symposium on Non-Perturbative and Symmetry
Methods in Field Theory
 (June 22--24, 2006, Budapest, Hungary).
The full collection is available at
\href{http://www.emis.de/journals/SIGMA/LOR2006.html}{http://www.emis.de/journals/SIGMA/LOR2006.html}}}

% Names of the authors for the title of the paper
\Author{Benjamin DOYON} 
\AuthorNameForHeading{B. Doyon}

\Address{Rudolf Peierls Centre for Theoretical Physics, Oxford University,\\
1 Keble Road, Oxford OX1 3NP, U.K.}

\Email{\href{mailto:b.doyon1@physics.ox.ac.uk}{b.doyon1@physics.ox.ac.uk}}

\URLaddress{\url{www-thphys.physics.ox.ac.uk/user/BenjaminDoyon}}

\ArticleDates{Received October 09, 2006, in f\/inal form December
07, 2006; Published online January 11, 2007}

\Abstract{We review the concept of f\/inite-temperature form
factor that was introduced recently by the author in the context
of the Majorana theory. Finite-temperature form factors can be
used to obtain spectral decompositions of f\/inite-temperature
correlation functions in a way that mimics the form-factor
expansion of the zero temperature case. We develop the concept in
the general factorised scattering set-up of integrable quantum
f\/ield theory, list certain expected properties and present the
full construction in the case of the massive Majorana theory,
including how it can be applied to the calculation of correlation
functions in the quantum Ising model. In particular, we include
the ``twisted construction'', which was not developed before and
which is essential for the application to the quantum Ising
model.}

\Keywords{f\/inite temperature; integrable quantum f\/ield theory;
form factors; Ising model}

\Classification{81T40}

\def\ba#1{\begin{array}{#1}\displaystyle}
\newcommand{\ea}{\end{array}}

\newcommand{\no}{\nonumber}
\newcommand{\n}{\nonumber\\}
\def\mato#1{\left(\ba{#1}} % exemple: \mato{cc} a & b \\ c & d \matf
\def\matf{\ea\right)}

\def\sect#1{\section{#1}}
\def\ssect#1{\subsection{#1}}
\def\sssect#1{\subsubsection{#1}}

\def\lt#1{\left#1}
\def\rt#1{\right#1}
\def\t#1{\tilde{#1}}
\def\h#1{\hat{#1}}
\def\b#1{\bar{#1}}
\def\frc#1#2{\frac{#1}{#2}}
\newcommand{\G}{\Gamma}
\newcommand{\p}{\partial}
\newcommand{\prin}{\underline{\mathrm{P}}}
\newcommand{\Pexp}{{\cal P}\exp}
\newcommand{\vac}{{\rm vac}}
\newcommand{\bra}{\langle}
\newcommand{\ket}{\rangle}
\newcommand{\Z}{{\mathbb{Z}}}
\newcommand{\N}{{\mathbb{N}}}
\newcommand{\R}{{\mathbb{R}}}
\newcommand{\C}{{\mathbb{C}}}

\newcommand{\Or}{{\cal O}}

\newcommand{\si}{\sigma}
\newcommand{\ep}{\epsilon}
\newcommand{\varep}{\varepsilon}

\newcommand{\Tr}{{\rm Tr}}

\newcommand{\Res}{{\rm Res}}
\newcommand{\End}{{\rm End}}

\newcommand{\ft}{{\cal L}}
\newcommand{\braL}{\langle\langle}
\newcommand{\ketL}{\rangle\rangle_\beta}
\newcommand{\vacft}{{\rm vac}_{\cal Y}}
\newcommand{\al}{\alpha}
\newcommand{\rx}{{\rm x}}
\newcommand{\om}{\omega}
\newcommand{\Evac}{{\cal E}}
\newcommand{\Zft}{{\cal Z}}
\newcommand{\mft}{{\Omega}}
\newcommand{\nft}{{{}^*_*}}
\newcommand{\nh}{{:\hspace{-0.1mm}}}
\newcommand{\meft}{\Upsilon}
\def\le#1{{\,}^{#1}\hspace{-0.7mm}}

\section{Introduction}

Relativistic quantum f\/ield theory (QFT) at f\/inite temperature
is a subject of great interest which has been studied from many
viewpoints (see, for instance, \cite{Kapusta}). An important task
when studying a model of QFT is the calculation of correlation
functions of local f\/ields, which are related to local
observables of the underlying physical model. For instance,
two-point correlation functions are related to response functions,
which can be measured and which provide precise information about
the dynamics of the physical system at thermodynamic equilibrium.
Although applications to particle physics often can be taken to be
at zero temperature, many applications to condensed matter require
the knowledge of the ef\/fect of a non-zero temperature on
correlation functions.

In this article, we review and develop further the ideas of
\cite{I} for studying f\/inite-temperature correlation functions
in integrable quantum f\/ield theory.

In recent years, thanks to advances in experimental techniques
allowing the identif\/ication and study of quasi-one-dimensional
systems (see for instance \cite{BourbonnaisJ99,Gruner}), there has
been an increased interest in calculating correlation functions in
1+1-dimensional integrable models of QFT (for applications of
integrable models to condensed matter systems, see for instance
the recent review~\cite{EsslerK04}). Integrable models are of
particular interest, because in many cases, the spectrum of the
Hamiltonian in the quantization on the line is known exactly (that
is, the set of particle types and their masses), and most
importantly, matrix elements of local f\/ields in eigenstates of
the Hamiltonian, or form factors, can be evaluated exactly by
solving an appropriate Riemann-Hilbert problem in the rapidity
space \cite{VergelesG76,Weisz77,KarowskiW78,BergKW79,Smirnov}.

At zero temperature, correlation functions are vacuum expectation
values in the Hilbert space of quantization on the line. The
knowledge of the spectrum on the line and the matrix elements of
local f\/ields then provides a useful expansion of two-point
functions at space-like distances, using the resolution of the
identity in terms of a basis of common eigenstates of the momentum
operator and of the Hamiltonian. This is a useful representation
because it is a large-distance expansion, which is hardly
accessible by perturbation theory, and which is often the region
of interest in condensed matter applications. Form factor
expansions in integrable models at zero temperature have proven to
provide a good numerical accuracy for evaluating correlation
functions in a wide range of energies, and combined with conformal
perturbation theory give correlation functions at all energy
scales (an early work on this is \cite{ZamolodchikovAl91}).

One would like to have such an ef\/f\/icient method for
correlation functions at f\/inite (non-zero) temperature as well.
Two natural (mathematically sound) ways present themselves:
\begin{itemize}\itemsep=0pt
\item ``Form factor'' expansion in the quantization on the circle.
It is a general result of QFT at f\/inite temperature
\cite{Matsubara55,Kubo57,MartinS59} that correlation functions, at
space-like distances, can be evaluated by calculating correlation
functions of the same model in space-time with Euclidean (f\/lat)
metric and with the geometry of a cylinder, the ``imaginary time''
wrapping around the cylinder whose circumference is the inverse
temperature. In this picture, one can quantize on the circle (that
is, taking space as being the circle, and Euclidean time the
line), and correlation functions become vacuum expectation values
in the Hilbert space of this quantization scheme. Then, one can
insert a resolution of the identity in terms of a~complete set of
states that diagonalise both the generator of time translations
and of space translations, as before, and one obtains a
large-distance expansion for f\/inite-temperature correlation
functions.

Unfortunately, the two ingredients required (the energy levels in
the quantization on the circle and the matrix elements of local
f\/ields) are not known in general in integrable quantum f\/ield
theory. We should mention, though, that exact methods exist to
obtain non-linear integral equations that def\/ine the energy
levels (from thermodynamic Bethe ansatz techniques, from
calculations \`a la Destri-de Vega and from the so-called BLZ
program), and that matrix elements of local f\/ields were studied,
for instance, in
\cite{Smirnov98a,Smirnov98b,ElburgS00,MussardoRS03}. Also, in the
Majorana theory, the spectrum is known (since this is a free
theory), and matrix elements of the primary ``interacting'' twist
f\/ields were calculated in \cite{Bugrij00,Bugrij01} from the
lattice Ising model, and in a simpler way in \cite{FonsecaZ01}
directly in the Majorana theory using the free-fermion equations
of motion and the ``doubling trick''. \item Spectral decomposition
on the space of ``f\/inite-temperature states''. The concept of
f\/inite-temperature states, interpreted as particle and hole
excitations above a ``thermal vacuum'', was initially proposed
more than thirty years ago and developed into a mature theory
under the name of thermo-f\/ield dynamics
\cite{LeplaeUM74,ArimitsuU87,ArimitsuU87a} (for a review, see for
instance \cite{Henning95}). Ideas of 1+1-dimensional integrable
quantum f\/ield theory were not applied to this theory until
recently. In \cite{AmaralB05}, the concept of bosonization in
thermo-f\/ield dynamics was studied, and, of most interest to the
present review, in \cite{I} the concept of f\/inite-temperature
form factor was developed -- matrix elements of local f\/ields on
the f\/inite-temperature Hilbert space. There, it was studied in
depth in the free Majorana theory, both for general free f\/ields
(f\/inite normal-ordered products of the free Majorana fermion
f\/ields -- including the energy f\/ield) and for twist f\/ields.
It was found that a Riemann--Hilbert problem again characterises
f\/inite-temperature form factors of twist f\/ields, but that this
Riemann--Hilbert problem presents important modif\/ications with
respect to the zero-temperature case. Solutions were written
explicitly for primary ``order'' and ``disorder'' twist f\/ields,
and the full f\/inite-temperature form factor expansions of
two-point functions were written and interpreted as Fredholm
determinants.
\end{itemize}

An interesting discovery of \cite{I} is that these two methods are
actually related: it is possible to evaluate form factors on the
circle from (analytical continuations of) the f\/inite-temperature
form factors, and the analytical structure of f\/inite-temperature
form factors (and of the measure involved in the expansion of
correlation functions) is directly related to the spectrum in the
quantization on the circle. This provided a new way of evaluating
form factors of twist f\/ields on the circle, and most
importantly, gave a clear prescription for the integration
contours in the f\/inite-temperature form factor expansion
(naively plagued with singularities). The requirements brought on
f\/inite-temperature form factors by this relation constitute, in
a way, a generalisation of the modularity requirements found in
conformal f\/ield theory for constructing correlation functions
from conformal blocks.

\looseness=1 It is important to realise, though, that both
expansions for correlation functions are not equivalent. The
f\/irst one gives an expansion at large (space-like) distances,
whereas the second can be used to obtain both large-distance and,
expectedly with more work, large-time expansions. Indeed, the
f\/inite-temperature form factor expansion can naturally be
deformed into an expansion in the quantization on the circle
through the relation mentioned above \cite{I}. It is expected that
it can also be manipulated to obtain large-time behaviours. A
manipulation of this type was done in \cite{AltshulerT05}. There,
going in reverse direction as what is described in~\cite{I}, the
expansion on the circle in the quantum Ising model was f\/irst
deformed into a kind of f\/inite-temperature form factor expansion
(without being recognised as such), which was then used to obtain
large-time dynamical correlation functions in a certain
``semi-classical'' regime (partly reproducing earlier results of
\cite{Sachdev96} and \cite{SachdevY97}). This manipulation,
however, neglected contributions that may change the asymptotic
behaviour, and a more complete derivation of the large-time
behaviours from f\/inite-temperature form factor expansions is
still missing. In particu\-lar, for the quantum Ising model, the
Fredholm determinant representation of \cite{I} and those obtained
in the present paper may be of use, following the technology
reviewed in \cite{KorepinBogoliubovIzergin93} (work is in
progress~\cite{DoyonGamsa06}).

It is worth noting that the method we review here is not adapted
to providing information about one-point functions at
f\/inite-temperature. Various works exist concerning such objects
\cite{Balog94,LeclairM99,Lukyanov01}. Work \cite{Balog94} is
interesting in that it uses the knowledge of the zero-temperature
form factors in order to deduce the f\/inite-temperature one-point
function of the energy f\/ield. The idea is to ``perform''
directly the f\/inite-temperature trace from the known matrix
elements. A~regu\-larisation is necessary, but the f\/inite-volume
one seems impossible to tackle. A certain convenient
regularisation was proposed there and shown to reproduce the known
f\/inite-temperature average energy. The idea of using this
regularisation for multi-point correlation functions has been
suggested and we are aware of results in this direction
\cite{EsslerKComm}, but it is not yet understood why in general
this should work.

Let us also mention that correlation functions of twist f\/ields
in the Majorana theory can be obtained as appropriate solutions to
non-linear dif\/ferential equations \cite{WuMTB76}. But at
f\/inite temperature, or on the geometry of the cylinder, these
equations are partial dif\/ferential equations in the coordinates
on the cylinder \cite{Perk80,Lisovyy02,FonsecaZ03}, and do not
immediately of\/fer a very useful tool for numerically evaluating
correlation functions, neither for analyzing their large-distance
and large-time behaviours.

The theory developed in \cite{I} for the Majorana case is still
incomplete. Twist f\/ields present certain complexif\/ications at
f\/inite temperature that are not present at zero temperature,
and, in order to describe all correlation functions, one also
needs a ``twisting'' of the construction of~\cite{I}, as it was
mentioned there. In addition, certain exponential pre-factors were
omitted in~\cite{I}. These two aspects are in fact essential for
applications of the results in the Majorana theory to correlation
functions in the quantum Ising model.

In this article we will review the ideas of \cite{I}, by
developing them in the general factorised scattering context of
integrable quantum f\/ield theory, and complete the work for the
Majorana theory. We will deduce many of the immediate properties
that arise in the general context for f\/inite-temperature form
factors, drawing on the ideas of \cite{I}, and we will present
both the untwisted and the twisted constructions. We will recall
the results for the Majorana theory, and extend them to the
twisted case, f\/inally giving the explicit representation for
correlation functions in the quantum Ising model.

The article is organised as follows. In Section \ref{review} we
review the form factor program at zero temperature, and in Section
\ref{ftcorrelation} we recall basic results about
f\/inite-temperature correlation functions. Then, in Section
\ref{ftprogram}, we describe the concept of f\/inite-temperature
states using the language of factorised scattering in integrable
QFT, we introduce the concept of f\/inite-temperature form factor
and we describe the resulting expansion of correlation function.
We also present the ideas underlying the relation between
f\/inite-temperature form factors and matrix elements in the
quantization on the circle, still in the general context. In
Section \ref{twistconstr}, we develop the basics of the twisted
construction. In Section \ref{formalstruct}, we present certain
formal results about the space of f\/inite-temperature states, and
in particular, we deduce a generalisation of the idea of ``mapping
to the cylinder'' that one uses in conformal f\/ield theory in
order to study correlation functions at f\/inite temperature
(again, this is a generalisation of ideas of~\cite{I}). Finally,
in Section \ref{ftMajorana}, we recall and extend the results
of~\cite{I} for the Majorana theory and its connection to the
quantum Ising model.

\renewcommand{\thefootnote}{\arabic{footnote}}
\setcounter{footnote}{0}

\section[Review of the zero-temperature form factor program
 in integrable quantum field theory]{Review of the zero-temperature
 form factor program\\ in integrable quantum f\/ield theory}
\label{review}

The Hilbert space of massive relativistic quantum f\/ield theory
is completely specif\/ied by f\/ixing the set $E$ of particle
types of the model. In 1+1 dimensions, every Hamiltonian
eigenstate is then described by choosing $k\in\N$ particle types
and by associating to them $k$ real numbers, the rapidities:
\[
    |\theta_1,\ldots,\theta_k\ket_{a_1,\ldots,a_k}
\]
with $a_i\in E$ and $\theta_i\in \R$ (and the order of the
rapidities/particle types is irrelevant -- hence a basis is
obtained by f\/ixing an ordering of the rapidities). The
Hamiltonian $H$ and the momentum $P$ act diagonally on these
states. In order to f\/ix their eigenvalues, one only has to f\/ix
the masses $m_a\in\R^+$ for every particle type $a\in E$. The
eigenvalues are then
\[
    H~:~E_k = \sum_{i=1}^k m_{a_i} \cosh\theta_i,\qquad
    P~:~p_k = \sum_{i=1}^k m_{a_i} \sinh\theta_i.
\]
Other symmetries of the model also act diagonally, and their
eigenvalues are f\/ixed by choosing charges associated to the
various particle types.

There are many possible bases of the Hilbert space, all described
as above. Two are of particular importance: the $in$ basis and the
$out$ basis. They describe, respectively, particles of the given
types and rapidities far in the past, and far in the future (in
non-integrable models, one should really include the additional
dependence on the impact parameters). The far past and the far
future are regions in time where all particles are so far apart
that they do not interact, and can be described as freely
propagating. The overlap between the $in$ basis and the $out$
basis gives the scattering matrix:
\[
    |\theta_1,\theta_2,\ldots\ket^{(in)}_{a_1,a_2,\ldots} = \sum_{a_1',a_2',\ldots} \int d\theta_1'd\theta_2'\cdots
    S_{a_1,a_2,\ldots}^{a_1',a_2',\ldots}(\theta_1,\theta_2,\ldots;\theta_1',\theta_2',\ldots)
    |\theta_1',\theta_2',\ldots\ket^{(out)}_{a_1',a_2',\ldots},
\]
where the number of particles in the $in$ state and in the $out$
states is generically dif\/ferent. The structure of the Hilbert
space and the Hamiltonian describe the particles and their
propagation, but it is the scattering matrix that encodes the
interaction, and in particular, the locality of relativistic
quantum f\/ield theory.

In integrable quantum f\/ield theory, the scattering matrix can be
determined from the physical requirements of unitarity and
crossing symmetry, from the integrability requirement of
factorisation and the lack of particle production, and from
minimality assumptions and the ``nuclear democracy'' (every pole
has a physical explanation through resonances from particles
already in the spectrum). All scattering processes can then be
described using only the two-particle scattering matrix $S_{a_1,
a_2}^{b_1, b_2}(\theta_1-\theta_2)$, $\theta_1>\theta_2$:
\[
    |\theta_1,\theta_2\ket^{(in)}_{a_1,a_2} = \sum_{b_1,b_2}
    S_{a_1,a_2}^{b_1,b_2}(\theta_1-\theta_2)
    |\theta_1,\theta_2\ket^{(out)}_{b_1,b_2}.
\]
It is convenient for this purpose to introduce the
Zamolodchikov--Faddeev algebra (from now on in this section,
summation over repeated indices will be implied)
\begin{gather}
    Z^{a_1}(\theta_1) Z^{a_2}(\theta_2) - S^{a_1, a_2}_{b_1,b_2}(\theta_1-\theta_2)
    Z^{b_2}(\theta_2) Z^{b_1}(\theta_1) = 0, \nonumber\\
    \b{Z}_{a_1}(\theta_1) \b{Z}_{a_2}(\theta_2) - S_{a_1, a_2}^{b_1,b_2}(\theta_1-\theta_2)
    \b{Z}_{b_2}(\theta_2) \b{Z}_{b_1}(\theta_1) = 0, \label{ZF} \\
    Z^{a_1}(\theta_1) \b{Z}_{a_2}(\theta_2) - S^{b_2,a_1}_{a_2,b_1}(\theta_2-\theta_1)
    \b{Z}_{b_2}(\theta_2) Z^{b_1}(\theta_1) = \delta^{a_1}_{a_2}\;\delta(\theta_1-\theta_2). \nonumber
\end{gather}
The $in$ basis and the $out$ basis are then two bases for the same
Fock space (actually, a generalisation of the concept of Fock
space) over this algebra, def\/ined simply by dif\/ferent ordering
of the rapidities:
\begin{gather*}
    Z^a(\theta)|\vac\ket = 0,\\
    |\theta_1,\ldots,\theta_k\ket_{a_1,\ldots,a_k}^{(in)} = \b{Z}_{a_1}(\theta_1) \cdots \b{Z}_{a_k}(\theta_k)
    |\vac\ket
    \qquad (\theta_1>\cdots>\theta_k),\\
    |\theta_1,\ldots,\theta_k\ket_{a_1,\ldots,a_k}^{(out)} = \b{Z}_{a_1}(\theta_1) \cdots \b{Z}_{a_k}(\theta_k)
    |\vac\ket
    \qquad (\theta_1<\cdots<\theta_k).
\end{gather*}
The natural Hermitian structure on this space gives
$(Z^a(\theta))^\dag = \b{Z}_a(\theta)$.

Once the Hilbert space has been identif\/ied with the Fock space
over the Zamolodchikov--Faddeev algebra, the algebra elements
$Z^a(\theta)$ and $\b{Z}_a(\theta)$ become operators with an
action on the Hilbert space. It turns out, from expected
properties of quantum f\/ield theory, that they induce very nice
properties on the objects (form factors)
\[
    F^{\Or}_{a_1,\ldots,a_k}(\theta_1,\ldots,\theta_k) \equiv \bra\vac| \Or(0,0) \b{Z}_{a_1}(\theta_1)\cdots
    \b{Z}_{a_k}(\theta_k)|\vac\ket,
\]
where $\Or(x,t)$ is a local f\/ield of the model. Indeed, these
objects, def\/ined here for real rapidities, actually are (by
analytical continuation) meromorphic functions of the rapidities.
They can be determined through a set of analyticity requirements
and through the recursive determination of the residues at the
poles (form factor equations) \cite{KarowskiW78,Smirnov}:
\begin{enumerate}\itemsep=0pt
\item Meromorphicity: as functions of the variable
$\theta_i-\theta_j$, for any $i,j\in\{1,\ldots,k\}$, they are
analytic inside $0<{\rm Im}(\theta_i-\theta_j)<2\pi$ except for
simple poles; \item Relativistic invariance:
\[
    F_{a_1,\ldots,a_k}^\Or(\theta_1+\beta,\ldots,\theta_k+\beta) =
    e^{s(\Or)\beta} F_{a_1,\ldots,a_k}^\Or(\theta_1,\ldots,\theta_k),
\]
where $s(\Or)$ is the spin of $\Or$; \item Generalized Watson's
theorem:
\begin{gather*}
    F_{a_1,\ldots,a_j,a_{j+1},\ldots,a_k}^\Or(\theta_1,\ldots,\theta_j,\theta_{j+1},\ldots,\theta_k)\\
    \qquad{} =
    S_{a_j,a_{j+1}}^{b_j,b_{j+1}}(\theta_j-\theta_{j+1})
    F_{a_1,\ldots,b_{j+1},b_j,\ldots,a_k}^\Or(\theta_1,\ldots,\theta_{j+1},\theta_{j},\ldots,\theta_k);
\end{gather*}
\item Locality:
\[
    F_{a_1,\ldots,a_{k-1},a_k}^\Or(\theta_1,\ldots,\theta_{k-1},\theta_k+2\pi i) =
    (-1)^{f_\Or f_\Psi} e^{2\pi i \om(\Or,\Psi)}
    F_{a_k,a_1,\ldots,a_{k-1}}^\Or(\theta_k,\theta_1,\ldots,\theta_{k-1}),
\]
where $f_\Or$ is 1 if $\Or$ is fermionic, 0 if it is bosonic,
$\Psi$ is the fundamental f\/ield associated to the particle
$a_k$, and $\om(\Or,\Psi)$ is the {\em semi-locality index} (or
mutual locality index) of $\Or$ with respect to $\Psi$ (it will be
def\/ined in Subsection \ref{semilocal}); \item Kinematic pole: as
function of the variable $\theta_n$, there are poles at
$\theta_j+i\pi$ for $j\!\in\!\{1,\ldots,k-1\}$, with residue
\begin{gather*}
     iF_{a_1,\ldots,a_k}^\Or(\theta_1,\ldots,\theta_k) \sim
    C_{a_k,b_j} \frc{F_{a_1,\ldots,\h{a}_j,\ldots,a_{k-1}}(\theta_1,\ldots,\h{\theta}_j,\ldots,\theta_{k-1})}{
      \theta_k-\theta_j-i\pi}  \\
    \qquad{} \times\Big( \delta_{a_1}^{b_1} \cdots \delta_{a_{j-1}}^{b_{j-1}}
        S_{a_{j+1},a_j}^{b_{j+1},c_j}(\theta_{j+1}-\theta_j)
    S_{a_{j+2},c_j}^{b_{j+2},c_{j+1}}(\theta_{j+2}-\theta_{j}) \cdots
        S_{a_{k-1},c_{k-3}}^{b_{k-1},b_j}(\theta_{k-1}-\theta_j) \\
   \qquad{}  - (-1)^{f_\Or f_\Psi} e^{2\pi i \om(\Or,\Psi)}
        \delta_{a_{k-1}}^{b_{k-1}} \cdots \delta_{a_{j+1}}^{b_{j+1}}
        S_{a_{j},a_{j-1}}^{c_{j},b_{j-1}}(\theta_{j}-\theta_{j-1})
    S_{c_j,a_{j-2}}^{c_{j-1},b_{j-2}}\\
    \qquad{}\times (\theta_{j}-\theta_{j-2}) \cdots
        S_{c_3,a_{1}}^{b_{j},b_1}(\theta_{j}-\theta_1) \Big),
\end{gather*}
where a hat means omission of the argument, and $C_{a_k,b_j}$ is the conjugation matrix. 

\item Bound-state
poles: there are additional poles in the strip $0 < {\rm
Im}(\theta_i-\theta_j) < \pi$ if bound states are present, and
these are the only poles in that strip.
\end{enumerate}
Form factors can in turn be used to obtain a large-distance
expansion of two-point correlation functions of local f\/ields:
\begin{gather*}
    \bra\vac|\Or_1(x,t)\Or_2(0,0)|\vac\ket =
    \sum_{k=0}^\infty\sum_{a_1,\ldots,a_k}
    \int\frc{d\theta_1\cdots d\theta_k}{k!}    e^{-it\sum_j E_j+ix\sum_j p_j}\\
\phantom{\bra\vac|\Or_1(x,t)\Or_2(0,0)|\vac\ket =}{}\times
    \bra\vac|\Or(0,0)|\theta_1,\ldots,\theta_k\ket_{a_1,\ldots,a_k}
    {\ }_{a_1,\ldots,a_k}
    \bra\theta_1,\ldots,\theta_k|\Or(0,0)|\vac\ket. \!
\end{gather*}
A large-distance expansion is ef\/fectively obtained by shifting
all rapidity variables by $\pi/2$ in the positive imaginary
direction, and by using relativistic invariance. This gives a
formula which looks as above, but with the replacement
$e^{-it\sum_j E_j+ix\sum_j p_j} \mapsto e^{-r\sum_j
m_{a_j}\cosh(\theta_j)}$ where $r=\sqrt{x^2-t^2}$. It turns out
that this is numerically extremely ef\/f\/icient in most
integrable models that were studied.

\section{Finite temperature correlation functions}
\label{ftcorrelation}
\subsection{Traces}

Physical correlation functions at f\/inite temperature are
obtained by taking a statistical average of quantum averages, with
Boltzmann weights $e^{-\beta E}$ where $E$ is the energy of the
quantum state and $\beta$ is the inverse temperature. They are
then represented by traces over the Hilbert space:
\begin{gather}\label{ftcorr}
    \braL \Or(x,t)\cdots\ketL = \frc{\Tr\lt[e^{-\beta H}
    \Or(x,t)\cdots\rt] }{\Tr\lt[e^{-\beta H}\rt]}.
\end{gather}
Since all matrix elements of local f\/ields are known in many
integrable models, it would seem appropriate to write the trace as
an explicit sum over all states of the Hilbert space, and to
introduce resolutions of the identity between operators inside the
trace, in order to evaluate f\/inite-temperature correlation
functions. However, this method does not account correctly for the
fact that at f\/inite temperature, states that contribute to the
trace are very far from the vacuum. Yet, it turned out to give
good results in the case of correlation functions with only one
operator~\cite{LeclairM99,Lukyanov01,Balog94}.

\subsection{Quantization on the circle}

On the other hand, traces as above can be represented by vacuum
expectation values on the Hilbert space ${\cal H}_\beta$ of
quantization on the circle of circumference $\beta$. Indeed, a
consequence of the imaginary-time formalism \cite{Matsubara55} is
the Kubo--Martin--Schwinger (KMS) identity
\cite{Kubo57,MartinS59},
\begin{gather}\label{KMS}
    \braL \Or(x,t) \cdots \ketL =
    (-1)^{f_\Or} \braL \Or(x,t-i\beta) \cdots \ketL,
\end{gather}
where $(-1)^{f_\Or}$ is a sign accounting for the statistics of
$\Or$ (it is negative for fermionic operators and positive for
bosinic operators), and where the dots ($\cdots$) represent local
f\/ields (that are also local with respect to $\Or$) at time $t$
and at positions dif\/ferent from $x$. Then, f\/inite-temperature
correlation functions can be written as
\begin{gather}\label{circle}
    \braL \Or(\tau,i{\rm x}) \cdots \ketL = \big(e^{i\pi s/2}\cdots\big)\,{\ }_\beta\bra\vac| \h\Or({\rm x},\tau)
    \cdots |\vac\ket_\beta,
\end{gather}
where $s$ is the spin of $\Or$, and there are factors $e^{-i\pi
s/2}$ for all operators in the correlation function. The operator
$\h\Or(\rx,\tau)$ is the corresponding operator acting on the
Hilbert space ${\cal H}_\beta$ of quantization on the circle, with
space variable $\rx$ (parameterizing the circle of circumference
$\beta$) and Euclidean time variable $\tau$ (on the line). The
vector $|\vac\ket_\beta$ is the vacuum in this Hilbert space.
Below, we will mostly be interested in fermionic models, that is,
models with a ``fundamental'' fermion f\/ield (which creates from
the vacuum single-particle states). For such models, one can think
of at least two sectors in the quantization on the circle:
Neveu--Schwartz~(NS) and Ramond~(R), where the fundamental fermion
f\/ields are anti-periodic and periodic, respectively, around the
circle. The trace (\ref{ftcorr}) with insertion of operators that
are local with respect to the fermion f\/ields naturally
corresponds to the NS sector due to the KMS identity. This is the
sector with the lowest vacuum energy.

The representation (\ref{circle}) immediately leads to a
large-distance expansion of f\/inite-temperature correlation
functions, through insertion of the resolution of the identity on
the Hilbert space ${\cal H}_\beta$:
\begin{gather}
    {}_\beta\bra\vac_1|\h\Or(\rx,\tau)\h\Or(0,0)|\vac_2\ket_\beta =  \sum_{k=0}^\infty
    \sum_{n_1,\ldots,n_k} \frc{e^{\sum_j n_j \frc{2\pi i \rx}\beta + (\Delta\Evac- E_{n_1,\ldots,n_k})
    \tau}}{k!} \label{circleffexp}\\
    \phantom{{}_\beta\bra\vac_1|\h\Or(\rx,\tau)\h\Or(0,0)|\vac_2\ket_\beta = }{}\times
    {\ }_\beta\bra\vac_1|\h\Or(0,0)|n_1,\ldots,n_k\ket_\beta
    {\ }_\beta\bra n_1,\ldots,n_k|\h\Or(0,0)|\vac_2\ket_\beta,\nonumber
\end{gather}
where the eigenstates of the momentum operator and of the
Hamiltonian on the circle are parametrized by discrete variables
$n_j$'s. The vacua $|\vac_1\ket_\beta$ and $|\vac_2\ket_\beta$ may
be in dif\/ferent sectors, and these sectors may be dif\/ferent
than the sector where the excited states
$|n_1,\ldots,n_k\ket_\beta$ lie (this situation occurs when
considering semi-local operators as is recalled in the
Subsection~\ref{semilocal} below). The quantity $\Delta\Evac$ is
the dif\/ference between the vacuum energies of the vacuum state
$|\vac_1\ket_\beta$ and of the vacuum above which the states
$|n_1,\ldots,n_k\ket_\beta$ are constructed. The states
$|n_1,\ldots,n_k\ket_\beta$ and the excitation energies $E_{n_1,\ldots,n_k}$
may also depend on additional discrete parameters (quantum
numbers, particle types), on which one has to sum as well.

This form is valid for any integrable model on the circle.
However, this Hilbert space has a very complicated structure, even
in integrable quantum f\/ield theory; for instance the energy
levels $E_{n_1,\ldots,n_k}$ are not known in closed form. Also,
there is yet no known procedure in general integrable quantum
f\/ield theory for evaluating form factors on this Hilbert space.
Moreover, this representation does not provide large-time (real
time) expansions, since it inherently gives f\/inite-temperature
correlation functions in Euclidean time.

\subsection[Semi-locality: $U(1)$ twist fields]{Semi-locality: $\boldsymbol{U(1)}$ twist f\/ields}
\label{semilocal}

If the model we are considering has internal global symmetries,
then there are local twist f\/ields associated to them. Twist
f\/ields are of interest, because they usually correspond to some
order parameter. We will clarify the correspondence between
order/disorder parameters in the quantum Ising chain and twist
f\/ields in the Section~\ref{twoptfct}. The f\/irst appearances of
certain twist f\/ields in the context of the Ising statistical
model can be found in \cite{KadanoffC71,SchroerT78}, but we are
going to describe twist f\/ields here in more general terms (see,
for instance, the lecture notes~\cite{DoyonLectures05}).

Twist f\/ields are not local with respect to the fundamental
f\/ields associated to a given particle (but are with respect to
the energy density). If the symmetry to which they are associated
is $U(1)$ or a subgroup of it (and if the fundamental f\/ield
transform by multiplication by a phase), then the twist f\/ields
are said to be semi-local with respect to the fundamental f\/ield.
In the quantization scheme on the line, a twist f\/ield, which we
will generically denote by $\sigma$, gives rise to a pair of
operators, which we will denote by $\si_+(x,t)$ and $\si_-(x,t)$,
having a cut towards the right (positive $x$ direction) and
towards the left (negative $x$ direction), respectively. These
operators lead to the same correlation functions at zero
temperature.

When considering correlation functions at f\/inite temperature,
things are more subtle. The exact shape of the cuts are
unimportant, but it is important if the cut is towards the right
or towards the left. This is because the insertion of an operator
$\si_\pm(x,t)$ that is semi-local with respect to the fundamental
f\/ield $\Psi(x,t)$ may af\/fect the vacuum sector in the
correspondence to expectation values in the quantization on the
circle. Semi-locality can be expressed through the exchange
relations
\begin{gather}\label{semilocp}
    \Psi(x,t)\si_+(x',t) = (-1)^{f_\Psi f_\si} e^{-2\pi i \omega \Theta(x-x')} \si_+(x',t) \Psi(x,t) \qquad(x\neq x')
\end{gather}
and
\begin{gather}\label{semilocm}
    \Psi(x,t)\si_-(x',t) = (-1)^{f_\Psi f_\si} e^{2\pi i \omega \Theta(x'-x)} \si_-(x',t) \Psi(x,t)\qquad(x\neq x'),
\end{gather}
where $\Theta(x)$ is Heaviside's step function and $\om$ is the
semi-locality index associated to the pair $(\Psi,\si)$. Taking
here the fundamental f\/ield to be fermionic (because this is what
will be of interest in the following -- the case of bosonic
fundamental f\/ields is straightforward to work out), it is
a~simple matter to generalise the KMS identity to (using
$f_\Psi=1$)
\begin{gather*} %\label{genKMS}
    \braL \Psi(x,t) \si_+(x',t) \cdots \ketL = \lt\{\ba{ll}
    -e^{-2\pi i \omega} \braL \Psi(x,t-i\beta) \si_+(x',t) \cdots \ketL & (x\to\infty),\vspace{1mm} \\
    -\braL \Psi(x,t-i\beta) \si_+(x',t) \cdots \ketL & (x\to-\infty), \ea\rt. \\
    \braL \Psi(x,t) \si_-(x',t) \cdots \ketL = \lt\{\ba{ll}
    - \braL \Psi(x,t-i\beta) \si_-(x',t) \cdots \ketL & (x\to\infty),\vspace{1mm} \\
    -e^{2\pi i \omega} \braL \Psi(x,t-i\beta) \si_-(x',t) \cdots \ketL & (x\to-\infty), \ea\rt.
\end{gather*}
where the dots ($\cdots$) represent f\/ields that are local with
respect to the fermion f\/ield $\Psi$, at time~$t$ and at
positions dif\/ferent from $x$. Then, in the correspondence of the
trace with a vacuum expectation value in the quantization on the
circle, one of the vacua will be in a dif\/ferent sector, in
accordance with these quasi-periodicity relations. Denoting by
$|\vac_{\nu}\ket_\beta$ the vacuum in the quantization on the
circle with quasi-periodicity condition $\Psi\mapsto e^{-2\pi i
\nu} \Psi$ around the circle in the positive space ($\rx$)
direction, we have
\begin{gather}\label{circletwistp}
    \braL \si_+(\tau,i\rx) \cdots \ketL = \big(e^{i\pi s/2}\cdots\big)\,{\ }_\beta\bra\vac_{\frc12+\omega}|
    \h\si(\rx,\tau) \cdots |\vac_{{\rm NS}}\ket_\beta
\end{gather}
and
\begin{gather}\label{circletwistm}
    \braL \si_-(\tau ,i\rx) \cdots  \ketL = \big(e^{i\pi s/2}\cdots\big)\,{\ }_\beta\bra\vac_{{\rm NS}}|
    \h\si(\rx,\tau) \cdots |\vac_{\frc12-\omega}\ket_\beta,
\end{gather}
where $|\vac_{{\rm NS}}\ket_\beta = |\vac_{\frc12}\ket_\beta$ is
the NS vacuum, and where the dots ($\cdots$) represent operators
that are local with respect to the fundamental fermion f\/ields.

With many insertions of semi-local operators, similar phenomena
arise. This change of the vacuum sector has an important ef\/fect:
under translation in the $x$ direction, the insertion of an
operator $\si_\pm(x,t)$ inside a trace produces an additional real
exponential factor, due to the dif\/ference between the vacuum
energies of the dif\/ferent sectors; that is, the trace is not
translation invariant. It is convenient to represent this lack of
translation invariance of traces, in the case where many
semi-local operators are multiplied, by considering ``modif\/ied''
transformation properties of this product of semi-local operators.
Consider the product $\si_{\eta_1}^{\om_1}(x_1,t_1) \cdots
\si_{\eta_k}^{\om_k}(x_k,t_k)$ where $\eta_i=\pm$ and we have
indicated explicitly the semi-locality indices $\om_i$. Then,
inside traces at temperature $\beta$ with insertion of operators
that are local with respect to the fundamental fermion f\/ield, we
have
\begin{gather}
    e^{-iP\delta} \si_{\eta_1}^{\om_1}(x_1,t_1) \cdots \si_{\eta_k}^{\om_k}(x_k,t_k)
    e^{iP\delta} = e^{-\Delta \Evac\;\delta} \;\si_{\eta_1}^{\om_1}(x_1+\delta,t_1)
    \cdots \si_{\eta_k}^{\om_k}(x_k+\delta,t_k), \nonumber\\
    \Delta \Evac = \Evac\lt[\frc12 + \sum_{i=1\atop \eta_i=+}^k \omega_i\rt]-
    \Evac\lt[\frc12 - \sum_{i=1\atop \eta_i=-}^k \omega_i\rt], \label{DeltaE}
\end{gather}
where $\Evac[\nu]$ is the energy of the vacuum
$|\vac_\nu\ket_\beta$.

\section{A space of ``f\/inite-temperature states''\\ in integrable quantum f\/ield theory}
\label{ftprogram}

In \cite{I}, it was suggested that the dif\/f\/iculties in
obtaining large-distance or large-time expansions of
f\/inite-temperature correlation functions can be overcome by
constructing a f\/inite-temperature Hilbert space in terms of
objects with nice analytic structure, in analogy with the
zero-tem\-pe\-ra\-tu\-re case. The program was carried out
explicitly in the free massive Majorana theory (considering, in
particular, ``interacting'' twist f\/ields). As we said in the
introduction, the idea of a f\/inite-temperature Hilbert space is
far from new, but it is in \cite{I} that it was f\/irst developed
in the context of an integrable quantum f\/ield theory.

\subsection{General idea}

The idea of the construction is simply to consider the space $\ft$
of endomorphisms of ${\cal H}$ as a~Hilbert space with inner
product structure
\[
    (A,B) = \frc{\Tr\lt(e^{-\beta H} A^\dag B\rt)}{\Tr\lt(e^{-\beta H} \rt)}.
\]
This Hilbert space is known as the Liouville space
\cite{vanHove86}. Note that
\[
    (A,B)^* = (B,A).
\]
There is then a (generically) one-to-two mapping from $\End(\cal
H)$ to $\End(\ft)$: to each operator $C$ acting on ${\cal H}$,
there are two operators, $\phi_L(C)$ and $\phi_R(C)$, acting on
$\ft$, def\/ined respectively by left action and by right action
of $C$ as follows:
\[
    (A,\phi_L(C) B) = \frc{\Tr\lt(e^{-\beta H} A^\dag C B\rt)}{\Tr\lt(e^{-\beta H} \rt)},\qquad
    (A,\phi_R(C) B) = \frc{\Tr\lt(e^{-\beta H} A^\dag B C\rt)}{\Tr\lt(e^{-\beta H} \rt)}.
\]
In particular, if $Q$ is a generator of a symmetry transformation
on ${\cal H}$, then $\phi_L(Q)-\phi_R(Q)$ is the generator on
$\ft$. The set of all operators on $\ft$ that are in the image of
at least one of $\phi_L$ or~$\phi_R$ will be denoted
$\End_{LR}(\ft)$.

The main power of this construction, from our viewpoint, is the
possibility to obtain large-distance or large-time expansions at
f\/inite temperature, in analogy with the zero-temperature case,
using a resolution of the identity on the space $\ft$. Indeed,
suppose we have a complete set of orthonormal operators
$D(\theta_1,\ldots,\theta_k),\;\theta_1> \cdots>\theta_k\in\R,
\;k\in\N$:
\[
    (D(\theta_1,\ldots,\theta_k),D(\theta_1',\ldots,\theta_l')) = \delta_{k,l}\;
    \delta(\theta_1-\theta_1')\cdots\delta(\theta_k-\theta_k').
\]
Then, we can decompose any inner product as a sum of products of
inner products:
\[
    (A,B) = \sum_{k=0}^\infty \int_{\theta_1>\cdots>\theta_k}
    d\theta_1\cdots d\theta_k\; (A,D(\theta_1,\ldots,\theta_k))\,
    (D(\theta_1,\ldots,\theta_k),B).
\]
This is a non-trivial relation equating a trace on the left-hand
side to a sum of products of traces on the right-hand side.

\subsection{A natural basis}

A natural complete set of operators in integrable quantum f\/ield
theory can be obtained as follows (more precisely, one should
consider an appropriate completion of the set below). First,
def\/ine a larger set of particle types $E_\ft= E\oplus E$, the
elements being the couples $\alpha = (a,\ep)$ for $a\in E$ and
$\ep=\pm$. For notational convenience, def\/ine
$\b{Z}_\alpha=\b{Z}_a$ if $\ep=+$ and $\b{Z}_\alpha=Z^a$ if
$\ep=-$. Then, we have a complete set with
\[
    D_{\al_1,\ldots,\al_k}(\theta_1,\ldots,\theta_k) =
    \b{Z}_{\al_1}(\theta_1)\cdots \b{Z}_{\al_k}(\theta_k)
\]
for
\begin{gather}\label{ordering}
    \al_i\in E_\ft,\qquad \theta_1>\cdots>\theta_k \in\R,\qquad k\in\N.
\end{gather}
In fact, it will be convenient to def\/ine the operators
$D_{\al_1,\ldots,\al_k}(\theta_1,\ldots,\theta_k)$ for any
ordering of the rapidities, and to def\/ine them as being exactly
zero when two rapidities collide (in order to avoid overlap with
operators with smaller $k$):
\begin{gather}\label{ops}
    D_{\al_1,\ldots,\al_k}(\theta_1,\ldots,\theta_k) = \lt\{\ba{ll}
    \b{Z}_{\al_1}(\theta_1)\cdots \b{Z}_{\al_k}(\theta_k),& \theta_i\neq\theta_j\quad \forall~i\neq j,\vspace{1mm}\\
    0, & \theta_i=\theta_j \quad \mbox{for some $i\neq j$}.\ea\rt.
\end{gather}

These operators will form a very useful set if the matrix elements
(with $\Or(x,t)$ a local operator and $\Or^\dag(x,t)$ its
Hermitian conjugate on $\cal H$)
\[
    (\Or^\dag(x,t),D_{\al_1,\ldots,\al_k}(\theta_1,\ldots,\theta_k))
\]
have simple analytical properties; for instance, if the value of
this function for a certain ordering of the rapidities is the one
obtained by analytical continuation from that for a dif\/ferent
ordering. Then, it may be possible to write down equations similar
to the zero-temperature form factor equations (what we will call
the ``f\/inite-temperature form factor equations'') and to solve
them. Two clues suggest that this may be so, at least for models
with diagonal scattering (see below): f\/irst, these objects
specialise to the zero-temperature form factors when the
temperature is sent to zero and all signs $\ep_i$'s are set to
$+$, and second, in \cite{I} the f\/inite-temperature form factor
equations were indeed written and solved in the free massive
Majorana theory (this will be recalled in Section
\ref{ftMajorana}).

Although the operators (\ref{ops}) form a complete set, they are
not orthonormal. In the case of diagonal scattering (and this is
the only case we will consider from now on)
\[
    S_{a_1,a_2}^{b_1,b_2}(\theta) = \delta_{a_1}^{b_1} \delta_{a_2}^{b_2} S_{a_1,a_2}(\theta)
\]
(without summation over repeated indices), it is possible to write
down all inner products in a simple way, using the
Zamolodchikov--Faddeev algebra (\ref{ZF}) and the cyclic property
of the trace:
\begin{gather}\label{inner}
    (D_{\al_1,\ldots,\al_k}(\theta_1,\ldots,\theta_k),
    D_{\al_1',\ldots,\al_l'}(\theta_1',\ldots,\theta_l'))
    = \delta_{k,l} \prod_{i=1}^k\frc{\ep_i^{1-f_{a_i}} \delta_{a_i,a_i'} \delta(\theta_i-\theta_i')}{
      1-(-1)^{f_{a_i}}\,e^{-\ep_i \beta m_{a_i}\cosh\theta_i}},
\end{gather}
where from unitarity, $(-1)^{f_a} \equiv S_{a,a}(0)=\pm1$ (this
corresponds to the statistics of the particle of type $a$, as an
asymptotically free particle). Here, we have assumed the ordering
(\ref{ordering}) for both members of the inner product.

Note that simple ``crossing'' relations hold for operators in
$\End_{RL}(\ft)$:
\begin{gather} \label{crossingL}
    (D_{\al_1,\ldots,\al_k}(\theta_1,\ldots,\theta_k)\;,\;
    \phi_L(A) D_{\al_1',\ldots,\al_l'}(\theta_1',\ldots,\theta_l'))
     \\
   = e^{\ep_l'\beta m_{a_l'} \cosh\theta_l'}
    (D_{\al_1,\ldots,\al_k,\b\al_l'}(\theta_1,\ldots,\theta_k,\theta_l')\;,\;
    \phi_L(A) D_{\al_1',\ldots,\al_{l-1}'}(\theta_1',\ldots,\theta_{l-1}'))
    \qquad (\theta_l' \neq \theta_i ~\forall\, i)\nonumber
\end{gather}
and
\begin{gather} \label{crossingR}
    (D_{\al_1,\ldots,\al_k}(\theta_1,\ldots,\theta_k),
    \phi_R(A) D_{\al_1',\ldots,\al_l'}(\theta_1',\ldots,\theta_l'))
     \\  \qquad
    =(D_{\b\al_1',\al_1,\ldots,\al_k}(\theta_1',\theta_1,\ldots,\theta_k),
    \phi_R(A) D_{\al_2',\ldots,\al_{l}'}(\theta_2',\ldots,\theta_{l}')) \no
    \qquad (\theta_1' \neq \theta_i ~\forall\, i),\nonumber
\end{gather}
where $\overline{(a,\ep)} = (a,-\ep)$.

\subsection{Finite-temperature form factor expansion}

Inverting (\ref{inner}), and using the fact that the operators
(\ref{ops}) are eigenoperators of both the Hamiltonian and the
momentum operator, we get a spectral decomposition for two-point
functions $\braL\Or_1(x,t)\Or_2(0,0)\ketL$ (f\/inite-temperature
form factor expansion). In order to simplify the discussion, we
assume that $x>0$. This can always be achieved by taking complex
conjugation if necessary: $\braL\Or_1(x,t)\Or_2(0,0)\ketL^*
=\braL\Or_2^\dag(0,0)\Or_1^\dag(x,t)\ketL$ (a slightly dif\/ferent
formulation holds with $x<0$). We have
\begin{gather}\label{ftffexp}
    \braL\Or_1(x,t)\Or_2(0,0)\ketL  \\ {}= e^{\Delta \Evac\;x}
        \;\sum_{k=0}^\infty \sum_{\al_1,\ldots,\al_k}
    \int_{\{{\rm Im}(\theta_j)=\ep_j 0^+\}} \frc{d\theta_1\cdots
    d\theta_k\prod\limits_{j=1}^k\ep_j^{1-f_{a_j}}
    \;e^{\sum_{j=1}^k\ep_j (im_{a_j}x\sinh\theta_j-im_{a_j} t\cosh\theta_j)}}{
    k!\prod\limits_{j=1}^k\lt(1-(-1)^{f_{a_j}}e^{-\ep_j
    \beta m_{a_j} \cosh\theta_j}\rt)}\nonumber\\
     \phantom{{}={}}{} \times
    F^{\Or_1}_{\al_1,\ldots,\al_k}(\theta_1,\ldots,\theta_k;\beta)
    F^{\Or_2}_{-\al_k,\ldots,-\al_1}(\theta_k,\ldots,\theta_1;\beta),
    \nonumber
\end{gather}
where we have def\/ined {\em finite-temperature form factors} as
the normalised matrix elements\footnote{Note that by def\/inition
of the basis of states in $\ft$, the function
$F^\Or_{\al_1,\ldots,\al_n}(\theta_1,\ldots,\theta_n;\beta)$ has
no delta-function contributions at colliding rapidities.}:
\begin{gather} \label{ftff}
    F^\Or_{\al_1,\ldots,\al_k}(\theta_1,\ldots,\theta_k;\beta)
     \\ \qquad\qquad {}=\prod_{i=1}^n\lt[\ep_i^{1-f_{a_i}}
      \lt(1-(-1)^{f_{a_i}}\,e^{-\ep_i \beta m_{a_i}\cosh\theta_i}\rt)\rt]\;
    (\Or^\dag(0,0)\;,\;D_{\al_1,\ldots,\al_k}(\theta_1,\ldots,\theta_k)). \nonumber
\end{gather}
This normalisation is for later convenience (one may call it the
``free f\/ield'' normalisation). It leads in particular to the
following identity, which we have used to write the expansion:
\[
    \big(F^{\Or_2^\dag}_{\al_1,\ldots,\al_k}(\theta_1^*,\ldots,\theta_k^*;\beta)\big)^*
    = F^{\Or_2}_{-\al_k,\ldots,-\al_1}(\theta_k,\ldots,\theta_1;\beta)
\]
which essentially follows from (\ref{crossingLd}) below. In the
expansion (\ref{ftffexp}), we have symmetrised over the orderings
of rapidities. The quantity $\Delta \Evac$ is non-zero whenever
$\Or_1$ is a twist f\/ield $\sigma_{\eta_1}^{\om_1}$, and is given
by
\[
    \Delta \Evac = \lt\{ \ba{ll} \displaystyle \Evac\lt[\frc12 + \omega_1\rt] - \Evac\lt[\frc12 \rt] & (\eta_1=+), \vspace{1mm}\\
  \displaystyle  \Evac\lt[\frc12 \rt] - \Evac\lt[\frc12 - \omega_1\rt] & (\eta_1=-), \ea \rt.
\]
where $\Evac[\nu]$ is the energy of the vacuum
$|\vac_\nu\ket_\beta$ (see the discussion around (\ref{DeltaE})).

Some comments are due:
\begin{itemize}\itemsep=0pt
\item As we said, the factor $e^{\Delta \Evac\;x}$ is present
whenever the operator $\Or_1$ is semi-local, $\Or_1=\si_\eta^\om$,
with respect to the fundamental fermion f\/ield. It is as in
(\ref{DeltaE}) with $k=1$, and with $\omega_1=\om$ and
$\eta_1=\eta$ the semi-locality index and cut direction,
respectively, of $\Or_1$. It occurs because the operators
$\b{Z}_\al(\theta)$ can be expressed through integrals of the
fundamental fermion f\/ield. To be more precise, in order to
deduce it from the discussion around (\ref{DeltaE}), one has to
assume that although these integrals extend to $\pm\infty$ in the
$x$ direction, they only produce excited states, without changing
the sector. The presence of this exponential factor can in fact be
shown for the f\/inite-temperature form factors of the order
f\/ield $\si_{\pm}^{\frc12}$ in the Majorana theory. Indeed, as we
said, in the Majorana theory the traces
$F^{\si_{\pm}^{\frc12}}_{\al_1,\ldots,\al_n}(\theta_1,\ldots,\theta_n;\beta)$
were shown in \cite{I} to satisfy a set of recursive relations
which ultimately relate them to the one-point function (in the
case of the order f\/ield). Slightly generalising the derivation
to include an $x$ and $t$ dependence, this accounts for the phase
factors above. On the other hand, the one-point function of a
twist f\/ield is not translation invariant, as is clear
from~(\ref{circletwistp}) and (\ref{circletwistm}), the
transformation property being as in (\ref{DeltaE}). This is what
accounts for the real exponential factor (this factor was missing
in \cite{I}, because the one-point function was considered
translation invariant). \item When both $\Or_1$ and $\Or_2$ are
semi-local with respect to the fundamental fermion f\/ields, the
f\/inite-temperature form factor expansion (\ref{ftffexp}) is
valid only when the cut of $\Or_1$ extends towards the right
(positive $x$ direction) and that of $\Or_2$ extends towards the
left (negative $x$ direction). This will be justif\/ied in the
Subsection~\ref{ftcircle}. Note that with this prescription on the
directions of the cuts, one produces the correlation functions
\[
    {\ }_\beta\bra \vac_{\frc12+\om_1}| \h\Or_1(\rx,\tau) \h\Or_2(0,0) | \vac_{\frc12 - \om_2}\ket_\beta
\]
when written in the quantization on the circle, with $x=\tau$ and
$t=i\rx$. This is a restriction, as not all vacua on ${\cal H}_\beta$
can be obtained. For instance, one would like to evaluate
correlation functions of twist f\/ields with the NS vacuum. This
restriction will be lifted in Section~\ref{twistconstr}. \item Had
we not put a small imaginary part to the rapidities in the
integrals, the expansion~(\ref{ftffexp}) would have been plagued
by singularities: as the rapidity associated to a particle of type
$(a,\ep)$ becomes equal to that associated to a particle of type
$(a,-\ep)$, poles are expected to appear in the
f\/inite-temperature form factors (kinematic poles). This
expectation comes from the intuition from zero-temperature form
factors, and from the fact that these singularities indeed occur
in the f\/inite-temperature form factors of twist f\/ields in the
Majorana theory, as was calculated
 in~\cite{I}. There it was shown that a proper solution is obtained by slightly deforming the integration
contours as indicated above:
\begin{gather}\label{presc}
    {\rm Im}(\theta_j) = \ep_j 0^+.
\end{gather}
That this is the right prescription still in the interacting case
will be argued in Subsection~\ref{ftcircle}. \item It is important
to realise that the expansion (\ref{ftffexp}) is not directly a
large-distance or a large-time expansion. But it can be made so as
follows (for large-time expansions, this requires some work).
First, with further displacement of the integration contours in
the directions of (\ref{presc}), more precisely with ${\rm
Im}(\theta_j) = \ep_j \,\pi/2$, the expansion (\ref{ftffexp})
becomes an expansion at large $x^2-t^2$ (recall that we consider
$x>0$ for simplicity). In order to perform this contour
displacement, one needs to know about the analytical structure of
the integrands; this will be brief\/ly discussed in
Subsection~\ref{ftcircle}. Second, the integrals involved in
(\ref{ftffexp}) can be made convergent in time-like regions
$t^2-x^2>0$ by deforming the contours in the following way:
\begin{gather*}
    |t|>|x|,~t>0:\quad  {\rm Im}(\theta_j) = -{\rm sgn}({\rm Re}(\theta_j))\,\ep_j 0^+, \\
    |t|>|x|,~t<0:\quad {\rm Im}(\theta_j) = {\rm sgn}({\rm Re}(\theta_j))\,\ep_j 0^+.
\end{gather*}
These deformations necessitate the addition of residues coming
from the kinematic poles. These residues will lead to powers of
the time variable, which will need to be re-summed. Note that it
was assumed in \cite{AltshulerT05} that considering the
contributions near the singularities at colliding rapidities, the
expansion gives the leading  in some semi-classical region, which
should include a large-time limit $t^2-x^2\to\infty$. The full
contour deformation should give a def\/inite answer as to the
large-time dynamics (work is in progress \cite{DoyonGamsa06}).
\item When calculating the spectral density (from the Fourier
transform of the two-point function), the expansion
(\ref{ftffexp}) does produce an expansion with terms of lesser and
lesser importance as the particle number is increased, at least
for large enough energies. However, one does not have the
situation where the spectral density is known exactly up to a
certain energy depending on the number of particles considered, as
happens at zero temperature. It would be very interesting to have
a full analysis of the spectral density at f\/inite temperature.
\end{itemize}

\subsection[From finite-temperature states to the quantization on the circle]{From f\/inite-temperature states to the quantization on the circle}
\label{ftcircle}

A great part of the structure of the f\/inite-temperature form
factor expansion can be understood according to the following
idea.

Suppose that we have a model of quantum f\/ield theory; more
precisely, let us consider a~statistical f\/ield theory, on a
space with Euclidean signature. Let us quantize it with a certain
choice of space $x$ of inf\/inite extent, and Euclidean time
$t^E$. If we were starting from a Lorentzian quantum f\/ield
theory, with real time $t$, we would just be considering the Wick
rotated variable $t=-it^E$. Then, the Hilbert space is the space
of f\/ield conf\/igurations on $x$, with appropriate asymptotic
conditions. On this space, we choose a vacuum $|\vac\ket$ such
that correlation functions are vacuum expectation values. Now
suppose that a basis of states is chosen such that the generator
of $x$ translations is diagonalised. The operator producing $x$
translations is unitary, of the form $e^{-iPx}$, where $P$ is the
Hermitian generator. Suppose that the states are parametrised by
the real eigenvalues $p$ of the operator $P$. Since space is of
inf\/inite extent, then $p$ takes all real values.

Then, formally, if we were to ``analytically continue the theory''
towards positive imaginary eigenvalues $p=iE$, the operator
producing $x$ translations would have the form $e^{Hx}$ for some
Hermitian $H$ with eigenvalues $E$. The claim is that the operator
$H$ is still the generator of $x$ translation, but now in a
dif\/ferent quantization scheme (that is, on a dif\/ferent Hilbert
space), where $x=\tau$ is the Euclidean time and $t^E = -\rx$ is
the space variable. Indeed, in that quantization scheme, the
operator producing Euclidean time translations is $e^{H\tau}$ with
$H$ the Hamiltonian (generator of time translations).

This formal analytical continuation has to be made more precise.
Consider matrix elements of local operators $\bra\vac|
\Or(x,t^E)|p\ket$ with states $|p\ket$ of $P$-eigenvalue $p$.
Then, this matrix element has singularities as function of $p$ on
the positive imaginary axis, and the positions of these
singularities exactly coincide with the eigenvalues of the
Hamiltonian $H$ in the quantization scheme where $x$ is Euclidean
time. Moreover, the analytical continuation of the matrix element
towards these singularities gives the matrix element of the same
operator in the quantization scheme where $x$ is Euclidean time.
In relativistic quantum f\/ield theory, the singularities are
branch cuts coming from the measure $\prod_i\sqrt{p_i^2+m_i^2}$
(with $p=\sum_ip_i$), and the statement about analytical
continuation is just crossing symmetry. This claim was also
verif\/ied to be true in the free Dirac theory on the Poincar\'e
disk \cite{Doyon03}, where the singularities are poles and the
residues must be taken.

In the case of present interest, our claim is that the
``analytical continuation'' of the Hilbert space $\ft$ is nothing
else than the Hilbert space ${\cal H}_\beta$ of quantization on
the circle. This was verif\/ied explicitly in \cite{I} in the free
Majorana theory.

This claim is made relatively clear by comparing the
f\/inite-temperature form factor expansion~(\ref{ftffexp}) and the
expansion in the quantization scheme on the
circle~(\ref{circleffexp}), which must agree. The analytical
continuation we talked about is obtained by shifting the contours
of the rapidities $\theta_j$ by the amount $\ep_j\,i\pi/2$: then
the exponential factors of (\ref{ftffexp}) and of
(\ref{circleffexp}) indeed agree, under the identif\/ication
$x=\tau$, $t=i\rx$. This displacement of the contours can be
performed while keeping all integrals convergent: we impose $x>0$,
$|x|>|t|$ (space-like region) keeping $x$ and $t$ f\/ixed, and
make only at the end the analytical continuation $t=i\rx$. Hence,
we see that keeping the integrals convergent leads to keeping the
operators time-ordered in the quantization on the circle. Note
that it is here that the condition $x>0$ becomes important: the
analytical conditions that def\/ine the f\/inite-temperature form
factors will be seen below as consequence of this analytical
continuation, and depend, at least for twist f\/ields, on our
choice of sign of $x$. A dif\/ferent choice of sign would have
required a shift in a dif\/ferent direction, and would have
imposed dif\/ferent conditions on the f\/inite-temperature form
factors.

Of course, the series themselves must agree, but it is natural to
assume that they agree individually for each term with a f\/ixed
number of excitations (at least this can be expected for
integrable models, where there is no particle production -- see
section \ref{formalstruct} for a discussion). Then, these terms
will agree for all local f\/ields if the following conditions are
satisf\/ied:
\begin{itemize}\itemsep=0pt
\item The factor
\[
 \frc{F^{\Or_1}_{\al_1,\ldots,\al_k}(\theta_1,\ldots,\theta_k;\beta)
    F^{\Or_2}_{-\al_k,\ldots,-\al_1}(\theta_k,\ldots,\theta_1;\beta)}{
    \prod_{j=1}^k\lt(1-(-1)^{f_{a_j}}e^{-\ep_j
    \beta m_{a_j} \cosh\theta_j}\rt)}
\]
can be written as
\[
    \rho_{\al_1,\ldots,\al_k}(\theta_1,\ldots,\theta_k)~
    \t{F}^{\Or_1}_{\al_1,\ldots,\al_k}(\theta_1,\ldots,\theta_k;\beta)
    \t{F}^{\Or_2}_{-\al_k,\ldots,-\al_1}(\theta_k,\ldots,\theta_1;\beta),
\]
where
$\t{F}^{\Or}_{\al_1,\ldots,\al_k}(\theta_1,\ldots,\theta_k;\beta)$
does not have a pole in the strip ${\rm Im}(\theta_j) \in
[0,\ep_j\pi/2]$ for all local f\/ields $\Or$, and the measure
$\rho_{\al_1,\ldots,\al_k}(\theta_1,\ldots,\theta_k)$ has poles at
$\theta_j = q_j + \ep_j\,i\pi/2$ for various real~$q_j$ (more
precisely, there are sets $s_l^{(k)}=\{q_1,\ldots,q_k\}$ for which
the measure $\rho$ has poles at such positions, choosing an order:
for instance, the pole as function of $\theta_1$, whose residue
has a pole as function of $\theta_2$, etc.). \item We have
\[
    \sum_{j=1}^k m_{a_j}\cosh q_j = E_{n_1,\ldots,n_k},\qquad
    \sum_{j=1}^k m_{a_j}\sinh q_j = \sum_{j=1}^{k} \frc{2\pi n_j}\beta,
\]
where the sets $\{q_j\}$ are the sets $s_l^{(k)}$ for the choice
$\ep_1=\dots=\ep_k=+$, and are are in one-to-one correspondence
with the possible conf\/igurations of numbers $\{n_1,\ldots,n_k\}$
(there may be ambiguities in this correspondence). The values
$\sum\limits_{j=1}^{k} \frc{2\pi n_j}\beta$ must be in $\Z$ or in
$\Z+\frc12$ (or, in general, $\Z+\om$), in order to implement the
correct quasi-periodicity conditions; this is where our
f\/inite-temperature form factor expansion (\ref{ftffexp}) is made
to agree with the KMS identity. \item The quantities
$\t{F}^{\Or_1}_{\al_1,\ldots,\al_k}(\theta_1,\ldots,\theta_k;\beta)$
for $\theta_j = q_j + i\pi/2$ and for $\ep_1=\dots=\ep_k=+$, times
the square root of the residue of
$\rho_{\al_1,\ldots,\al_k}(\theta_1,\ldots,\theta_k)$ at these
same values, are proportional to the matrix elements
${}_\beta\bra\vac|\Or|n_1,\ldots,n_k\ket_\beta$.
\end{itemize}

Note that for the case of one particle, one should recover the
energy spectrum $E_{n,a} = m_a\cosh q_n$ with $m_{a}\sinh q_n =
\frc{2\pi n}\beta$ and $n\in\Z$ if $f_a=0$, $n\in\Z+\frc12$ if
$f_a=1$. This indicates that
\[
    \rho_\al(\theta) =     \frc{1}{1-(-1)^{f_{a}}e^{-\ep
    \beta m_{a} \cosh\theta}}
\]
and that $\t{F}_{\al}(\theta) = F_\al(\theta)$.

Note also that in the free Majorana theory, one simply has
\[
    \rho_{\al_1,\ldots,\al_k}(\theta_1,\ldots,\theta_k) =
    \frc{1}{
    \prod\limits_{j=1}^k\lt(1+e^{-\ep_j
    \beta m \cosh\theta_j}\rt)}
\]
with $\t{F}=F$ for all excitation numbers, and this indeed
reproduces the right energy levels in the quantization on the
circle as well as the correct matrix elements~\cite{I}. In fact,
for any free theory we have $\t{F}=F$.

It is now possible to understand the prescription (\ref{presc})
for deforming the integration contours in order to avoid possible
kinematic poles in the f\/inite-temperature form factors. Indeed,
by the principles above, the f\/inite-temperature form factor
expansion is really an analytical continuation of the sum
representation of two-point functions coming from the quantization
on the circle. Hence, it is natural that integration contours be
def\/ined in the complex plane to avoid kinematic poles, and the
direction of the deformation is exactly the one giving the proper
correspondence between the expansions (\ref{ftffexp}) and
(\ref{circleffexp}).

It is also possible to understand the restrictions on the
directions of the cut of semi-local operators, as explained in the
second comment after (\ref{ftffexp}). Indeed, the
f\/inite-temperature form factor
$F^{\Or_1}_{\al_1,\ldots,\al_k}(\theta_1,\ldots,\theta_k;\beta)$
is, in a sense, the analytical continuation of the matrix element
$\bra\vac|\Or_1|n_1,\ldots,n_k\ket_\beta$ in the quantization on
the circle, which describes ``one half'' of the two-point
function. In a path integral formulation, this matrix element
corresponds to a path integral on the half-cylinder, say,
$\tau>0$, $\rx\in [0,\beta]$, $\beta\equiv 0$ with some boundary
condition at $\tau=0$ (the excited state) and some asymptotic
condition at $\tau\to\infty$ (the vacuum). But since by
construction the function
$F^{\Or_1}_{\al_1,\ldots,\al_k}(\theta_1,\ldots,\theta_k;\beta)$
has no ``knowledge'' of the other operator $\Or_2$ of the
two-point function, it always ``stand'' in the natural sector
given by the trace, even if this sector is modif\/ied by the cut
emanating from $\Or_2$ in the actual correlation function. Hence,
it is important that the cut associated to $\Or_2$ does not change
this sector, that is, that it does not af\/fect the conditions at
$\tau\to\infty$ neither those in the region of $\tau$ present
between the position of $\Or_1$ and that of $\Or_2$. Similar
arguments apply to the function
$F^{\Or_2}_{-\al_k,\ldots,-\al_1}(\theta_k,\ldots,\theta_1;\beta)$,
and this shows that the cut of $\Or_1$ must be towards the right,
and that of $\Or_2$, towards the left, when both are operators
associated to twist f\/ields.

To be more precise, if the cut associated to $\Or_2$ does af\/fect
the sector in which $\Or_1$ stands, then the only way to provide
this information is by modifying the choice of the discrete values
of rapidities in
$F^{\Or_1}_{\al_1,\ldots,\al_k}(\theta_1,\ldots,\theta_k;\beta)$
that will form the states on the circle; that is, to modify the
analytic structure of the measure $\rho$. This is indeed what is
expected to happen: f\/inite-temperature form factors of twist
f\/ields should have an analytical structure that provides the
appropriate shift of the poles of the measure in order to produce
the change of sector. Then, we actually expect this to be enough
information for the f\/inite-temperature form factor of $\Or_1$ to
be in this dif\/ferent sector whenever $\Or_1$ is local with
respect to the fundamental f\/ields corresponding to the particles
involved. In that case, the cut of $\Or_2$ can indeed be in any
direction, the expansion~(\ref{ftffexp}) will stay valid.
Otherwise, if $\Or_1$ is itself semi-local with respect to the
fundamental f\/ields, then a change of the analytic structure of
the measure is not enough, hence the cut of $\Or_2$ must be in
opposite direction. This phenomenon is indeed what is observed in
the Majorana theory.

We can circumvent the restriction on directions of the cuts of the
twist f\/ields by ``twisting'' the construction; this is done in
the next section.

\section{Twisted constructions}
\label{twistconstr}

\subsection{The twisted inner product}

The construction of the previous section can be fruitfully
modif\/ied when there is a $U(1)$ inva\-riance (or sub-group
thereof) in the theory, by changing the quasi-periodicity
properties of the fundamental fermion f\/ield in imaginary time.
We still consider the space $\ft$ of endomorphisms of ${\cal H}$,
but now as a Hilbert space with a dif\/ferent inner product
structure:
\begin{gather}\label{twistinner}
    (A,B)_\om = \frc{\Tr\lt(e^{-\beta H + 2\pi i \om Q} A^\dag B\rt)}{
    \Tr\lt(e^{-\beta H + 2\pi i \om Q} \rt)} \equiv \braL A^\dag B\ketL^\om,
\end{gather}
where $Q$ is the Hermitian conserved charge associated to the
$U(1)$ symmetry. Then we again have
\[
    (A,B)_\om^* = (B,A)_\om.
\]

Now, we can still consider, in order to have a basis, the set of
operators (\ref{ops}). We have
\[
    Z_\al(\theta) e^{2\pi i \om Q} = e^{-2\pi i \om q(\al)} e^{2\pi i \om Q} Z_\al(\theta),
\]
where $q(\al)$ is the charge\footnote{In particular the charge of
the fundamental fermion, with $\ep=+$, is 1, and in general we
have $q(a,+) = -q(a,-)$.} of the excitation $\al$. Again with
diagonal scattering, using this it is possible to write down all
inner products (\ref{twistinner}) of the operator (\ref{ops}) in a
simple way:
\begin{gather}
    (D_{\al_1,\ldots,\al_k}(\theta_1,\ldots,\theta_k),
    D_{\al_1',\ldots,\al_l'}(\theta_1',\ldots,\theta_l'))_\om\nonumber\\
   \qquad{} = \delta_{k,l} \prod_{i=1}^k\frc{\ep_i^{1-f_{a_i}} \delta_{a_i,a_i'} \delta(\theta_i-\theta_i')}{
      1-(-1)^{f_{a_i}}e^{2\pi i \om q(\al_i)} \,e^{-\ep_i \beta m_{a_i}\cosh\theta_i}} .\label{twinner}
\end{gather}
Here, we have assumed the ordering (\ref{ordering}) for both
members of the inner product.

Note that simple ``crossing'' relations hold again for operators
in $\End_{RL}(\ft)$:
\begin{gather}
    (D_{\al_1,\ldots,\al_k}(\theta_1,\ldots,\theta_k)\;,\;
    \phi_L(A) D_{\al_1',\ldots,\al_l'}(\theta_1',\ldots,\theta_l'))
    =     e^{-2\pi i \om q(\al_l')} e^{\ep_l'\beta m_{a_l'} \cosh\theta_l'}\nonumber\\
    \qquad{}\times
    (D_{\al_1,\ldots,\al_k,\b\al_l'}(\theta_1,\ldots,\theta_k,\theta_l'),
    \phi_L(A) D_{\al_1',\ldots,\al_{l-1}'}(\theta_1',\ldots,\theta_{l-1}'))
    \qquad (\theta_l' \neq \theta_i ~\forall\, i) \label{twcrossingL}
\end{gather}
and
\begin{gather}
    (D_{\al_1,\ldots,\al_k}(\theta_1,\ldots,\theta_k)\;,\;
    \phi_R(A) D_{\al_1',\ldots,\al_l'}(\theta_1',\ldots,\theta_l'))\nonumber\\
    \qquad{}  =
    (D_{\b\al_1',\al_1,\ldots,\al_k}(\theta_1',\theta_1,\ldots,\theta_k)\;,\;
    \phi_R(A) D_{\al_2',\ldots,\al_{l}'}(\theta_2',\ldots,\theta_{l}'))
    \qquad (\theta_1' \neq \theta_i ~\forall\, i).\label{twcrossingR}
\end{gather}

\subsection[Twisted finite-temperature form factor expansion]{Twisted f\/inite-temperature form factor expansion}

Inverting (\ref{twinner}), we now have the twisted
f\/inite-temperature form factor expansion (as before, we assume
that $x>0$):
\begin{gather}
    \braL \Or_1(x,t)\Or_2(0,0)\ketL^\om
       \nonumber\\
        \qquad {}=e^{\Delta \Evac\;x}
        \sum_{k=0}^\infty \sum_{\al_1,\ldots,\al_k}
    \int_{\{{\rm Im}(\theta_j)=\ep_j 0^+\}} \frc{d\theta_1\cdots
    d\theta_k\prod\limits_{j=1}^k\ep_j^{1-f_{a_j}} \;
    e^{\sum\limits_{j=1}^k\ep_j (im_{a_j}x\sinh\theta_j-im_{a_j} t\cosh\theta_j)}}{
    k!\prod\limits_{j=1}^k\lt(1-(-1)^{f_{a_j}}e^{2\pi i \om q(\al_j)}e^{-\ep_j
    \beta m_{a_j} \cosh\theta_j}\rt)}\nonumber\\
     \qquad\phantom{{}={}}{}\times
    \le{\om} F^{\Or_1}_{\al_1,\ldots,\al_k}(\theta_1,\ldots,\theta_k;\beta)
    \le{\om} F^{\Or_2}_{-\al_k,\ldots,-\al_1}(\theta_k,\ldots,\theta_1;\beta),\label{twftffexp}
\end{gather}
where we have def\/ined {\em twisted finite-temperature form
factors} as the normalised matrix elements\footnote{Again, note
that the function
$F^\Or_{\al_1,\ldots,\al_n}(\theta_1,\ldots,\theta_n;\beta)$ has
no delta-function contributions at colliding rapidities.}:
\begin{gather}
    \le\om F^\Or_{\al_1,\ldots,\al_n}(\theta_1,\ldots,\theta_n;\beta)\nonumber\\
  {}= \prod_{i=1}^n\lt[\ep_i^{1-f_{a_i}}
      \big(1-(-1)^{f_{a_i}}e^{2\pi i \om q(\al_i)}\,e^{-\ep_i \beta m_{a_i}\cosh\theta_i}\big)\rt]
    (\Or^\dag(0,0),D_{\al_1,\ldots,\al_n}(\theta_1,\ldots,\theta_n))_\om. \label{twftff}
\end{gather}
and used the identity
\begin{gather}\label{cctw}
    \big(\le\om F^{\Or_2^\dag}_{\al_1,\ldots,\al_k}(\theta_1^*,\ldots,\theta_k^*;\beta)\big)^*
    = \le\om F^{\Or_2}_{-\al_k,\ldots,-\al_1}(\theta_k,\ldots,\theta_1;\beta),
\end{gather}
which essentially follows from (\ref{crossingLd}) below. Also, we
have symmetrised over the orderings of rapidities. The quantity
$\Delta \Evac$ is non-zero whenever $\Or_1$ is a twist f\/ield
$\sigma_{\eta_1}^{\om_1}$, and is now given by
\[
    \Delta \Evac = \lt\{ \ba{ll} \displaystyle \Evac\lt[\frc12 + \om + \omega_1\rt] - \Evac\lt[\frc12 +\om\rt]
     & (\eta_1=+),\vspace{1mm}\\
   \displaystyle \Evac\lt[\frc12 +\om\rt] - \Evac\lt[\frc12 +\om - \omega_1\rt] & (\eta_1=-), \ea \rt.
\]
where $\Evac[\nu]$ is the energy of the vacuum
$|\vac_\nu\ket_\beta$ (see the discussion around (\ref{DeltaE})).
Again, when both $\Or_1$ and $\Or_2$ are semi-local with respect
to the fundamental fermion f\/ield, the f\/inite-temperature form
factor expansion (\ref{ftffexp}) is valid only when the cut of
$\Or_1$ extends towards the right (positive $x$ direction) and
that of $\Or_2$ extends towards the left (negative $x$ direction).
This is justif\/ied in the same way as before, through the
relation between (\ref{twftffexp}) and a form factor expansion on
the circle (\ref{circleffexp}). Note that with this prescription
on the directions of the cuts, one now produces the correlation
functions
\[
    {\ }_\beta\bra \vac_{\frc12+\om+\om_1}| \h\Or_1(\rx,\tau) \h\Or_2(0,0) | \vac_{\frc12 + \om - \om_2} \ket_\beta
\]
with $x=\tau$ and $t=i\rx$. With $\om_1=-\om_2$, it is now
possible to have the NS vacuum by choosing $\om=-\om_1$.

\section{Formal structure and a generalisation of CFT's mapping\\ to the cylinder}
\label{formalstruct}

\subsection[The space $\ft$ as a Fock space, and physical interpretation]{The space
 $\boldsymbol{\ft}$ as a Fock space, and physical interpretation}

We have seen how the space $\ft$ of operators on ${\cal H}$ can be
used to obtain inf\/inite series expressions of correlation
functions (\ref{ftffexp}) and (\ref{twftffexp}). The important
objects are the f\/inite-temperature form factors (\ref{ftff}), or
the twisted version (\ref{twftff}), which are certain matrix
elements on the space $\ft$. It will be convenient, here, to
introduce the normalised operators
\begin{gather}\label{normops}
    d_{\al_1,\ldots,\al_k}(\theta_1,\ldots,\theta_k) = \prod_{i=1}^k\frc{1}{g_{\al_i}(\theta_i)}\;
    D_{\al_1,\ldots,\al_k}(\theta_1,\ldots,\theta_k)
\end{gather}
with, in the general twisted case,
\[
    g_\al(\theta) = \frc{\ep^{1-f_a}}{
      1-(-1)^{f_{a}}\,e^{2\pi i \om q(\al)}\,e^{-\ep \beta m_{a}\cosh\theta}}.
\]
Then, we have
\[
    F^\Or_{\al_1,\ldots,\al_k}(\theta_1,\ldots,\theta_k;\beta)
    =
    (\Or^\dag(0,0)\;,\;d_{\al_1,\ldots,\al_k}(\theta_1,\ldots,\theta_k))
\]
and the ``crossing'' relations (\ref{crossingL}),
(\ref{crossingR}) change into
\begin{gather}
    (d_{\al_1,\ldots,\al_k}(\theta_1,\ldots,\theta_k)\;,\;
    \phi_L(A) d_{\al_1',\ldots,\al_l'}(\theta_1',\ldots,\theta_l'))_\om\nonumber\\
   \qquad{}=
    (d_{\al_1,\ldots,\al_k,\b\al_l'}(\theta_1,\ldots,\theta_k,\theta_l'),
    \phi_L(A) d_{\al_1',\ldots,\al_{l-1}'}(\theta_1',\ldots,\theta_{l-1}'))_\om
    \qquad (\theta_l' \neq \theta_i ~\forall\, i)\label{crossingLd}
\end{gather}
and
\begin{gather}
    (d_{\al_1,\ldots,\al_k}(\theta_1,\ldots,\theta_k),
    \phi_R(A) d_{\al_1',\ldots,\al_l'}(\theta_1',\ldots,\theta_l'))_\om
    =     e^{2\pi i \om q(\al_1')} e^{-\ep_1'\beta m_{a_1'} \cosh\theta_1'}\nonumber\\
    \qquad{}\times
    (d_{\b\al_1',\al_1,\ldots,\al_k}(\theta_1',\theta_1,\ldots,\theta_k)\;,\;
    \phi_R(A) d_{\al_2',\ldots,\al_{l}'}(\theta_2',\ldots,\theta_{l}'))_\om
    \qquad (\theta_1' \neq \theta_i ~\forall\, i).\label{crossingRd}
\end{gather}

In order to describe in a convenient way the space spanned by
$d_{\al_1,\ldots,\al_k}(\theta_1,\ldots,\theta_k)$, we introduce
the following operators acting on $\ft$:
\[
    \Zft_{\al}^\dag(\theta), \Zft_{\al}(\theta)  \in \End(\ft)
\]
with the following properties:
\begin{gather*}
    \Zft_\al^\dag(\theta) {\bf 1}_{{\cal H}} = d_{\al}(\theta),\qquad
    \Zft_{\al_1}^\dag(\theta_1) d_{\al_2,\ldots,\al_k}(\theta_2,\ldots,\theta_k) =
    d_{\al_1,\ldots,\al_k}(\theta_1,\ldots,\theta_k),\!\qquad
    \Zft_\al(\theta) {\bf 1}_{{\cal H}} = 0\!
\end{gather*}
and satisfying the following exchange relations:
\begin{gather}
    \Zft_{\al_1}(\theta_1) \Zft_{\al_2}(\theta_2) - S_{\al_1,\al_2}(\theta_1-\theta_2)
    \Zft_{\al_2}(\theta_2) \Zft_{\al_1}(\theta_1) = 0,\nonumber\\
    \Zft^\dag_{\al_1}(\theta_1) \Zft^\dag_{\al_2}(\theta_2) - S_{\al_1,\al_2}(\theta_1-\theta_2)
    \Zft^\dag_{\al_2}(\theta_2) \Zft^\dag_{\al_1}(\theta_1) = 0, \label{ZFft} \\
    \Zft_{\al_1}(\theta_1) \Zft^\dag_{\al_2}(\theta_2) - S_{\al_2,\al_1}(\theta_2-\theta_1)
    \Zft^\dag_{\al_2}(\theta_2) \Zft_{\al_1}(\theta_1) = \frc1{g_{\al_1}(\theta_1)}
    \delta_{\al_1,\al_2}\delta(\theta_1-\theta_2),\nonumber
\end{gather}
where $\delta_{\al_1,\al_2} = \delta_{a_1,a_2}
\delta_{\ep_1,\ep_2}$ and where
\[
    S_{\al_1,\al_2}(\theta) = \lt\{\ba{ll}
    S_{a_1,a_2}(\theta) & (\ep_1=\ep_2), \vspace{1mm}\\
    S_{a_2,a_1}(-\theta) & (\ep_1=-\ep_2).
    \ea \rt.
\]

The space ${\cal H}$ is then seen as a Fock space over the algebra
(\ref{ZFft}), with vacuum vector ${\bf 1}_{{\cal H}}$ annihilated
by $\Zft_\al(\theta)$. The algebra (\ref{ZFft}) has exactly the
structure of the Zamolodchikov--Faddeev algebra (\ref{ZF}) with
diagonal scattering, but with twice as many particles. The
physical interpretation is that the ``states''
$d_{\al_1,\ldots,\al_k}(\theta_1,\ldots,\theta_k)$ correspond to
conf\/igurations of stable ``additional particles'' ($\ep=+$) and
``missing particles'' or holes ($\ep=-$) in a thermal bath (we
will call both ``excitations''), both created by the operators
$\Zft^\dag_\al(\theta)$, and scattering through the matrix
$S_{\al_1,\al_2}(\theta_1-\theta_2)$. They are stable, since the
states with $n$ excitations have no overlap with those with
$n'\neq n$ excitations. It is both this stability and the fact
that matrix elements should have nice analytical properties that
suggests that the correspondence between expansions in the
quantization on the circle and f\/inite-temperature form factor
expansions holds individually for every term with a given
excitation number. Only in integrable quantum f\/ield theory can
we expect these two properties together.

In order to have a better picture of the ``particle'' and ``hole''
states that we are discussing, one should recall that we def\/ined
the $in$-states of ${\cal H}$ by multiple action of
$\b{Z}_a(\theta)$ on $|\vac\ket$ with ordered rapidities,
decreasing from left to right. When this order is not satisf\/ied,
one has an ``intermediate'' state, which is a useful concept only
in integrable quantum f\/ield theory. It corresponds to having
wave packets ordered such that some interact in the far past (like
for $out$ states), while others interact in the far future (like
for $in$ states). Essentially, the order of the operators
$\b{Z}_a(\theta)$ acting on $|\vac\ket$ corresponds to the order
of the wave packets themselves along the $x$ axis, when time is
taken to go ``upward'' (in the positive $y$ direction). When we
put an operator $\b{Z}_a(\theta)$ inside the f\/inite-temperature
trace (evaluating the trace in the $in$ basis, for instance), we
take situations with various numbers of $in$ particles at various
ordered rapidities, and put an additional wave packet far to the
left, generically producing ``intermediate'' states. This is the
sense in which the operator $\b{Z}_a(\theta)$ corresponds to an
additional particle in a thermal bath. Similarly, $Z_a(\theta)$ is
taking away a particle from the thermal bath, by f\/irst bringing
its wave packet far to the left. Recall that it is because we keep
the order of the wave packets f\/ixed while varying the rapidities
that matrix elements of operators on ${\cal H}$ in
``intermediate'' states are meromorphic functions. With the
previous discussion, this lends support to the fact that the basis
$D_{\al_1,\ldots,\al_k}(\theta_1,\ldots,\theta_k)$ of $\ft$ should
produce matrix elements with nice analytical properties.

It is also interesting to note a nice physical interpretation for
the expected kinematical poles in the f\/inite-temperature form
factors, occurring at colliding rapidities when they are
associated to opposite signs of $\ep$. These poles can be seen, in
the f\/inite-temperature form factor expansion, to lead to powers
of the time variable $t$, instead of exponential factors. This
corresponds to the fact that a particle and a hole can annihilate
and re-form themselves at arbitrary large time dif\/ferences
without cost in energy. For a given excitation number, various
powers in $t$ will occur, which can more or less be made in
correspondence with various simultaneous annihilating and
re-creating of particle-hole pairs. It it these processes that
make the computation of large-time dynamics from
f\/inite-temperature form factors dif\/f\/icult: a re-summation of
these powers of $t$ is necessary.

The fact that the Hilbert space $\ft$ is the same as an ordinary
Hilbert space with twice as many ``particles'' does not mean that
the f\/inite-temperature theory is the same as a zero-temperature
one with such particles. Indeed, another ingredient that def\/ines
a theory is the set of local opera\-tors (in particular, the
energy density), and this set looks very dif\/ferent on a
f\/inite-temperature Hilbert space. In fact, it would be very
interesting to study the structure of the energy density on~$\ft$.

\subsection{A tool for evaluating traces and mapping to the cylinder}

This is again a generalisation of a concept introduced in
\cite{I}. It is possible to construct the left action of
$\b{Z}_\al$ through the operators $\Zft^\dag_\al$ and
$\Zft_{\b\al}$ (the right action is more complicated, from the
def\/inition of the operators $\Zft^\dag_\al$ and $\Zft_\al$, but
it is suf\/f\/icient to consider the left action). Indeed, one can
verify that the mapping
\begin{gather}\label{phiL}
    \phi_L(\b{Z}_\al(\theta)) = \Zft^\dag_\al(\theta) g_\al(\theta) + \Zft_{\b\al}(\theta) g_{\b\al}(\theta)
    \qquad (\b\al = (a,-\ep) \mbox{ for } \al=(a,\ep))
\end{gather}
is an isomorphism of the algebra satisf\/ied by
$\b{Z}_\al(\theta)$:
\[
    \b{Z}_{\al_1}(\theta_1)\b{Z}_{\al_2}(\theta_2) - S_{\al_1,\al_2}(\theta_1-\theta_2)
    \b{Z}_{\al_2}(\theta_2)\b{Z}_{\al_1}(\theta_1) = \ep_2^{1-f_{a_2}} \delta_{\ep_1,-\ep_2} \delta_{a_1,a_2}
    \delta(\theta_1-\theta_2),
\]
using the property $S_{\b\al_2,\al_1}(\theta_2-\theta_1) =
S_{\al_1,\al_2}(\theta_1-\theta_2)$ as well as $g_\al(\theta) +
(-1)^{1-f_a} g_{\b\al}(\theta) = \ep^{1-f_a}$. From this one can
conclude the equality
\[
    ({\bf 1}_{{\cal H}},\phi_L(\b{Z}_{\al_1}(\theta_1)) \cdots \phi_L(\b{Z}_{\al_k}(\theta_k)) {\bf 1}_{{\cal H}})
    =
    ({\bf 1}_{{\cal H}},\b{Z}_{\al_1}(\theta_1) \cdots \b{Z}_{\al_k}(\theta_k))
\]
by bringing on both sides, for instance, all factors with $\ep=+$
to the right, by evaluating the resulting right-hand side
explicitly using the inner product (\ref{twinner}), and by
evaluating the left-hand side using the representation
(\ref{phiL}). Then, restrict all rapidities numbered
$j+1,\ldots,k$ to be dif\/ferent from one another, and restrict
all rapidities numbered $1,\ldots,j-1$ to be dif\/ferent from one
another. The left-hand side evaluates to
\[
    (\b{Z}_{\al_{j-1}}(\theta_{j-1})\cdots \b{Z}_{\al_{1}}(\theta_{1}),
    \phi_L(\b{Z}_{\al_j}(\theta_j))
    \b{Z}_{\al_{j+1}}(\theta_{j+1})\cdots \b{Z}_{\al_k}(\theta_k)),
\]
whereas the right-hand side evaluates to
\[
    (\b{Z}_{\al_{j-1}}(\theta_{j-1})\cdots \b{Z}_{\al_{1}}(\theta_{1}),
    \b{Z}_{\al_j}(\theta_j)
    \b{Z}_{\al_{j+1}}(\theta_{j+1})\cdots \b{Z}_{\al_k}(\theta_k)).
\]
The equality of these two expressions shows the equality for all
matrix elements, hence \linebreak shows~(\ref{phiL}).

From relation (\ref{phiL}), one can write explicitly products
$\b{Z}_{\al_1}(\theta_1) \cdots \b{Z}_{\al_k}(\theta_k)$ as linear
combinations of the basis elements
$d_{\al_1',\ldots,\al_l'}(\theta_1',\ldots,\theta_l')$ (for $l\leq
k$ and where primed variables form a~subset of un-primed
variables), using
\[
    \b{Z}_{\al_1}(\theta_1) \cdots \b{Z}_{\al_k}(\theta_k) =
    \phi_L(\b{Z}_{\al_1}(\theta_1)) \cdots \phi_L(\b{Z}_{\al_k}(\theta_k)) {\bf 1}_{{\cal H}}.
\]
In particular, using cyclic properties of the trace, one can relax
the restriction of having dif\/ferent rapidities in
(\ref{crossingLd}), and one obtains
\begin{gather}
    (d_{\al_1',\ldots,\al_l'}(\theta_1',\ldots,\theta_l'),
    \phi_L(A) \,d_{\al_1,\ldots,\al_k}(\theta_1,\ldots,\theta_k))\nonumber\\
    \qquad{}   =
    (d_{\al_1',\ldots,\al_{l-1}'}(\theta_1',\ldots,\theta_{l-1}'),
    \phi_L(A)\; d_{\al_1,\ldots,\al_{k},\al_l'}(\theta_1,\ldots,\theta_{k},\theta_l'))
    \nonumber\\
    \qquad{}+
    \sum_{j=1}^{k} S_{\al_j,\al_{j+1}}(\theta_j-\theta_{j+1}) \cdots S_{\al_j,\al_k}(\theta_j,\theta_k)
        \frc{\delta_{\al_j,\al_l'}\delta(\theta_j-\theta_l')}{g_{\al_j}(\theta_j)}\nonumber\\
        \qquad{}    \times (d_{\al_1',\ldots,\al_{l-1}'}(\theta_1',\ldots,\theta_{l-1}'),
    \phi_L(A) \,d_{\al_1,\ldots,\h\al_j,\ldots,\al_{k}}(\theta_1,\ldots,\h\theta_j,\ldots,\theta_{k})),
    \label{crossingLdgen}
\end{gather}
where the hat means that the variable is missing.

It is now possible to generalise, in some sense, the concept of
``mapping to the cylinder'' that can be used in conformal f\/ield
theory in order to evaluate f\/inite-temperature correlation
functions. In conformal f\/ield theory, one has a mapping (of
vertex operator algebras) $\Or\mapsto \h\Or$ such that correlation
functions of $\h\Or$ on the cylinder are equal to correlation
functions of $\Or$ on the plane (for instance, for the stress
energy tensor one has $T\mapsto \lt(\frc{\beta}{2\pi}
T+\frc{c}{24}\rt)z^{-2} $ where $w$ is the coordinate on the
cylinder, $z=e^{2\pi w/\beta}$ is the coordinate on the plane and
$c$ is the central charge). We can generalise this here at the
level of form factors. We seek a linear map $\mft$ from~$\ft$ to
$\ft$ such that
\begin{gather}
    ({\bf 1}_{\cal H},\phi_L(\mft A)
    d_{(a_1,+),\ldots,(a_k,+),(a_l',-),\ldots,(a_1',-)}(\theta_1,\ldots,\theta_k,\theta_l',\ldots,\theta_1'))
    \nonumber\\
    \qquad{}=
       {}_{a_1',\ldots,a_l'}\bra \theta_1',\ldots,\theta_l'| A | \theta_1,\ldots,\theta_k\ket_{a_1,\ldots,a_k}
    \qquad (\theta_i'\neq \theta_j\ \forall\, i,j),\label{condmft}
\end{gather}
where $A\in\ft$. This requirement is inspired by the fact that
\begin{gather*}
    \lim_{\beta\to\infty} ({\bf 1}_{\cal H},\phi_L(A)
    d_{(a_1,+),\ldots,(a_k,+),(a_l',-),\ldots,(a_1',-)}(\theta_1,\ldots,\theta_k,\theta_l',\ldots,\theta_1'))\\
    \qquad{} =
    {}_{a_1',\ldots,a_l'}\bra \theta_1',\ldots,\theta_l'| A | \theta_1,\ldots,\theta_k\ket_{a_1,\ldots,a_k}
    \qquad (\theta_i'\neq \theta_j\ \forall\, i,j)
\end{gather*}
thanks to (\ref{crossingLd}). Note that this in fact completely
f\/ixes the map $\phi_L\circ\mft$, thanks to
(\ref{crossingLdgen}), and by injection the map $\mft$, if it
exists.

In order to describe the map $\mft$, it is convenient to consider
elements of $\ft$ that have simple expectation values on ${\cal
H}$. We consider products where operators $\b{Z}_\al(\theta)$ are
normal-ordered with respect to the vacuum in ${\cal H}$: operators
$Z^a(\theta)$ are placed to the right of all operators
$\b{Z}_a(\theta)$, without taking any delta-function term (but
taking all $S$-matrices involved in the exchanges). We will denote
the normal-ordering of $A$ by the standard $\nh A\nh$. The set of
all normal-ordered operators spans $\ft$. On the other hand, there
is a natural normal-ordering with respect to the vacuum ${\bf
1}_{\cal H}$ in $\ft$: it is the one whereby operators $\Zft$ are
placed to the right of operators $\Zft^\dag$. We will denote the
normal-ordering of ${\cal A}\in \End(\ft)$ by $\nft {\cal A}\nft$.
Then, it is simple to see that for all normal-ordered $A$ (that
is, $\nh A\nh=A$),
\begin{gather}\label{rel0}
    \nft \phi_L(A) \nft
\end{gather}
can be put in place of $\phi_L(\mft(A))$ in (\ref{condmft}) in
order to have equality. Indeed if $A$ contains exactly $l$
operators of type $\b{Z}_a$ and $k$ operators of type $Z^a$, then
the equality is clear; otherwise, both sides are zero, hence the
equality still holds. However, this does not yet show that the map
$\mft$ exists.

Now, let us choose a basis in $\ft$, with normal-ordered elements
$A_i$ composed of products of f\/initely many operators
$\b{Z}_\al$. Certainly, the set $\phi_L(A_i)$ is not a basis in
$\End(\ft)$. However, if we are to project from the left with
${\bf 1}_{{\cal H}}$, then we do obtain a basis: $({\bf 1}_{{\cal
H}},\phi_L(A_i)\cdot)$ is a basis in the dual of $\ft$. Similarly,
the space of operators $\nft \phi_L(A_i)\nft$ gives another basis
in the dual of $\ft$ when projected from the left with ${\bf
1}_{{\cal H}}$. Hence, we have the change-of-basis relation
\begin{gather}\label{chgbasis}
    ({\bf 1}_{{\cal H}}, \nft\phi_L(A_i)\nft~\cdot) = \sum_{j=0}^{k_i} M_{i,j}~ ({\bf 1}_{{\cal H}}, \phi_L(A_j)
    \,\cdot),
\end{gather}
where the sum is f\/inite. Then we f\/ind, from the discussion
around (\ref{rel0}) and recalling that $\nh A_i\nh = A_i$,
\[
    \mft A_i = \sum_{j=0}^{k_i} M_{i,j} A_j.
\]

The change of basis (\ref{chgbasis}) could be calculated
explicitly, but there is a nice way of expressing it. Suppose we
can f\/ind an operator $\meft$ in $\End(\ft)$ such that
\[
    \meft {\bf 1}_{{\cal H}} = 0,\qquad [\meft,\Zft_\al(\theta)] = 0,\qquad [\meft,\Zft_\al^\dag(\theta)]
    g_{\al}(\theta)
    = \Zft_{\b\al}(\theta) g_{(a,-)}(\theta).
\]
Then one can verify that $e^{\meft} \nft \phi_L(\nh A \nh)\nft
e^{-\meft} {\bf 1}_{{\cal H}} = \nh A \nh$, hence that
\[
    \mft = e^{-\meft}.
\]
But using the algebra (\ref{ZFft}), one can see that
\begin{gather}\label{meft}
    \meft = \int d\theta \sum_a \Zft_{(a,-)}(\theta) \Zft_{(a,+)}(\theta) g_{(a,-)}(\theta)
\end{gather}
has the right properties. Hence we have found an explicit
expression for the map, acting on the space of operators on ${\cal
H}$, that transform f\/inite-temperature form factors into
zero-temperature form factors. This is the generalisation of the
concept of ``mapping to the cylinder'' in conformal f\/ield
theory. Finite-temperature form factors can then be calculated
using
\begin{gather*}
    ({\bf 1}_{\cal H},\phi_L(A)
    \,d_{(a_1,+),\ldots,(a_k,+),(a_l',-),\ldots,(a_1',-)}(\theta_1,\ldots,\theta_k,\theta_l',\ldots,\theta_1'))\\
    \qquad{} =
     {}_{a_1',\ldots,a_l'}\bra \theta_1',\ldots,\theta_l'| \lt(e^{\meft} A\rt)
    | \theta_1,\ldots,\theta_k\ket_{a_1,\ldots,a_k}.
\end{gather*}
It is important to note, however, that the operator $e^\meft$ does
not act on the initial Hilbert space~${\cal H}$, but rather on the
space of operators acting on it $\ft$. In the quantization on the
line, this is not isomorphic to ${\cal H}$. In conformal f\/ield
theory, one usually thinks about the quantization on the circle
around a f\/ixed point (radial quantization), and by the
operator-state correspondence, one then has an operator acting on
the Hilbert space that performs the mapping to the cylinder.

The action of $e^\meft$ on $A\in\ft$ can be made more explicit
using (\ref{meft}). Indeed, taking $A = \b{Z}_{\al_1}(\theta_1)
\cdots \b{Z}_{\al_k}(\theta_k)$, we can write
\begin{gather*}
    e^\meft (\b{Z}_{\al_1}(\theta_1) \cdots \b{Z}_{\al_k}(\theta_k))
    = e^\meft \phi_L(\b{Z}_{\al_1}(\theta_1)) \cdots \phi_L(\b{Z}_{\al_k}(\theta_k)) {\bf 1}_{{\cal H}}\\
 \phantom{e^\meft (\b{Z}_{\al_1}(\theta_1) \cdots \b{Z}_{\al_k}(\theta_k))}{}   = e^\meft \phi_L(\b{Z}_{\al_1}(\theta_1)) e^{-\meft} \cdots
        e^\meft \phi_L(\b{Z}_{\al_k}(\theta_k)) e^{-\meft} {\bf 1}_{{\cal H}}
\end{gather*}
and use
\begin{gather*}
    e^{\meft} \phi_L(\b{Z}_\al(\theta)) e^{-\meft} = \phi_L(\b{Z}_\al(\theta)) +
    \Zft_{\b\al}(\theta) g_{(a,-)}(\theta),\\
    \Zft_{\b\al}(\theta) \phi_L(\b{Z}_{\al'}(\theta')) = S_{\al,\al'}(\theta-\theta') \phi_L(\b{Z}_{\al'}(\theta'))
    \Zft_{\b\al}(\theta) + \delta_{\b{\al},\al'}\delta(\theta-\theta')
\end{gather*}
to bring all the $\Zft_\al(\theta)$ to the right, annihilating
${\bf 1}_{{\cal H}}$. This leads to a natural generalisation of
Wick's theorem, whereby $e^\meft A$ is written as $A$ + operators
where more and more contractions have been performed, the
contractions being given by
\[
    \mbox{contraction of $\b{Z}_\al(\theta)$ with $\b{Z}_{\al'}(\theta')$} = g_{(a,-)}(\theta)
    \delta_{\b\al,\al'}(\theta-\theta').
\]
If the $S$-matrix $S_{\al,\al'}(\theta-\theta')$ is equal to $\pm
1$ (free models), then this gives the standard Wick's theorem, and
in particular it can be applied to linear combinations of the type
$\int d\theta \b{Z}_{\al}(\theta) f_\theta$ as well. In free
models certain linear combinations of this type indeed represent
local f\/ields, and this immediately leads to the phenomenon of
``mixing'' that was described in \cite{I} (using slightly
dif\/ferent arguments). Of course, local f\/ields in interacting
models and twist f\/ields in general are not simply such linear
combinations, but rather are sums of operators with more and more
factors of $\b{Z}_\al(\theta)$ (since many-particle form factors
are non-zero). Hence all term will contribute to any given
f\/inite-temperature form factor, and it is a non-trivial matter
to re-sum these contributions.

\section{Results in the free massive Majorana theory}
\label{ftMajorana}

\subsection{Free massive Majorana fermions}

The free massive Majorana theory with mass $m$ can be described by
the action
\[
    {\cal A} = i\int d^2x (-\psi (\p_x+\p_t) \psi +  \b\psi (\p_x-\p_t) \b\psi - m\b\psi \psi).
\]
It is a model with only one particle, and with only $\Z_2$
internal symmetry, described by a change of sign of the fermion
f\/ields. In particular, the f\/ields $\psi$ and $\b\psi$ are both
real (hence the corresponding operators in any quantization scheme
are Hermitian). The quantization on the line is simple to
describe. Fermion operators are given by:
\begin{gather*}
    \psi(x,t) = \frc12\sqrt{\frc{m}{\pi}}\,\int d\theta\,
    e^{\theta/2} \lt(
    a(\theta)\,e^{ip_\theta x -i E_\theta t} +
    a^\dag(\theta)\,e^{-ip_\theta x + i E_\theta t}
    \rt),\\
    \b\psi(x,t) = -\frc{i}2\,\sqrt{\frc{m}\pi}\,\int
    d\theta\,
    e^{-\theta/2} \lt(
    a(\theta)\,e^{ip_\theta x -i E_\theta t} -
    a^\dag(\theta)\,e^{-ip_\theta x + iE_\theta t}    \rt),
\end{gather*}
where the mode operators $a(\theta)$ and their Hermitian conjugate
$a^\dag(\theta)$ satisfy the canonical anti-commutation relations
\begin{gather}\label{algtheta}
    \{a^\dag(\theta),a(\theta')\} = \delta(\theta-\theta')
\end{gather}
(other anti-commutators vanishing) and where
\[
    p_\theta = m\sinh\theta~,\quad E_\theta = m\cosh\theta.
\]
The fermion operators satisfy the equations of motion
\begin{gather}
    \b\p\psi(x,t) \equiv \frc12\lt(\p_x + \p_t\rt) \psi =
    \frc{m}2 \b\psi,\nonumber\\
    \p\b\psi(x,t) \equiv \frc12\lt(\p_x - \p_t\rt) \b\psi =
    \frc{m}2 \psi\label{edm}
\end{gather}
and obey the equal-time anti-commutation relations
\begin{gather}\label{equaltime}
    \{\psi(x,t),\psi(x',t)\} = \delta(x-x'),\qquad
    \{\b\psi(x,t),\b\psi(x',t)\} = \delta(x-x').
\end{gather}
The Hilbert space $\cal H$ is simply the Fock space over the
algebra (\ref{algtheta}) with vacuum vector $|\vac\ket$ def\/ined
by $a(\theta)|\vac\ket=0$. Vectors in $\cal H$ will be denoted by
\[
    |\theta_1,\ldots,\theta_k\ket = a^\dag(\theta_1)\cdots
    a^\dag(\theta_k)|\vac\ket.
\]
A basis is formed by taking, for instance,
$\theta_1>\cdots>\theta_k$. This is exactly the construction
described in Section~\ref{review}, with only one particle and
$S(\theta) = -1$. The Hamiltonian is given by
\[
    H = \int_{-\infty}^{\infty}
    d\theta\,m\cosh\theta\,a^\dag(\theta) a(\theta)
\]
and has the property of being bounded from below on $\cal H$ and
of generating time translations:
\begin{gather}\label{Hpsi}
    [H,\psi(x,t)] = -i\frc{\p}{\p t} \psi(x,t),
    \qquad [H,\bar\psi(x,t)] = -i \frc{\p}{\p t}
    \bar\psi(x,t).
\end{gather}

In the discussions of the previous sections, we also considered
quantization on the circle of circumference $\beta$. It will be
convenient to have the description of this quantization for the
present model, with anti-periodic (NS) conditions on the fermion
f\/ields. The fermion operators evolved in Euclidean time $\tau$
are:
\begin{gather*}
    \h\psi(\rx,\tau) = \frc1{\sqrt{2L}}
    \sum_{n\in\Z+\frc12}\frc{e^{\alpha_n/2}}{\sqrt{\cosh\alpha_n}}\,
    \lt(
    a_n\,e^{ip_n \rx - E_n\tau} +
    a^\dag_n\,e^{-ip_n \rx + E_n\tau}
    \rt),\\
    \h{\b\psi}(\rx,\tau) = -\frc{i}{\sqrt{2L}}
    \sum_{n\in\Z+\frc12}\frc{e^{-\alpha_n/2}}{\sqrt{\cosh\alpha_n}}\,
    \lt(
    a_n\,e^{ip_n \rx - E_n\tau} -
    a^\dag_n\,e^{-ip_n \rx + E_n\tau}
    \rt),
\end{gather*}
where the discrete mode operators $a_n$ and their Hermitian
conjugate $a^\dag_n$ satisfy the canonical anti-commutation
relations
\begin{gather}\label{algn}
    \{a^\dag_n,a_{n'}\} = \delta_{n,n'}
\end{gather}
(other anti-commutators vanishing) and where
\begin{gather}\label{alphan}
    p_n = m\sinh\alpha_n = \frc{2\pi n}L \qquad \left(n\in\Z+\frc12\right), \\
    E_n =    m\cosh\alpha_n.\nonumber
\end{gather}
The fermion operators satisfy the equations of motion (\ref{edm})
as well as the equal-time anti-commutation relations
(\ref{equaltime}) (with the replacement $\psi\mapsto\h\psi$ and
$\b\psi\mapsto\h{\b\psi}$); the latter is simple to derive from
the representation
\[
    \delta(x) = \frc1{L} \sum_{n\in\Z+\frc12}e^{ip_n x}
\]
of the delta-function, valid on the space of antiperiodic
functions on an interval of length $\beta$. The Hilbert space
${\cal H}_\beta$ is simply the Fock space over the algebra
(\ref{algn}) with vacuum vector $|\vac_{{\rm NS}}\ket_\beta$
def\/ined by $a_n|\vac_{{\rm NS}}\ket_\beta=0$. Vectors in ${\cal
H}_\beta$ will be denoted by
\[
    |n_1,\ldots,n_k\ket_\beta = a^\dag_{n_1}\cdots
    a^\dag_{n_k}|\vac_{{\rm NS}}\ket_\beta.
\]
A basis is formed by taking, for instance, $n_1>\cdots>n_k$. The
Hamiltonian (with vacuum energy) is given by
\[
    H_\beta = \Evac_{{\rm NS}} + \sum_{n\in\Z+\frc12} m\cosh\alpha_n\,a^\dag_n a_n
\]
and has the property of being bounded from below on ${\cal
H}_\beta$ and of generating time translations:
\[
    [H_\beta,\h\psi(\rx,\tau)] = \frc{\p}{\p \tau} \h\psi(\rx,\tau),
    \qquad [H_\beta,\h{\bar\psi}(\rx,\tau)] = \frc{\p}{\p \tau}
    \h{\bar\psi}(\rx,\tau)~.
\]
Our discussion was with the NS sector in mind, but it is not hard
to perform the quantization in the R sector. What will be
important for us are relative energies of the NS and R vacua:
\begin{gather}
    \Evac_{{\rm NS}} \equiv \Evac[1/2] =
    \varepsilon -
    \int_{-\infty}^{\infty} \frc{d\theta}{2\pi} \cosh\theta \ln\lt(1+e^{-m\beta\cosh\theta}\rt),\nonumber\\
    \Evac_{{\rm R}} \equiv \Evac[0] =  \varepsilon -
    \int_{-\infty}^{\infty} \frc{d\theta}{2\pi} \cosh\theta \ln\lt(1-e^{-m\beta\cosh\theta}\rt),
    \label{vacener}
\end{gather}
where we used the notation of the discussion around
(\ref{DeltaE}). Here, the vacuum energies of both sectors were
calculated in the same regularisation scheme and $\varepsilon$
contain terms that are common to both.

It is worth noting that the normalisation that we took is slightly
dif\/ferent from the more standard normalisation in conformal
f\/ield theory, that makes the f\/ields $\psi$ and $\b\psi$ not
real, but with def\/inite phase. With our normalisation, the
leading terms of the operator product expansions (OPE's)
$\psi(x,t)\psi(0,0)$ and $\bar\psi(x,t)\bar\psi(0,0)$ are given by
\begin{gather}\label{OPEpsipsi}
    \psi(x,t)\psi(0,0) \sim \frc{i}{2\pi(x-t)},
    \qquad
    \b\psi(x,t)\b\psi(0,0) \sim -
    \frc{i}{2\pi(x+t)}.
\end{gather}

\subsection[Twist fields]{Twist f\/ields}

Two f\/ields are of particular importance: they are two primary
twist f\/ields associated to the $\Z_2$ symmetry, which we will
denote by $\sigma$ and $\mu$ as is customary\footnote{In the
present section, the symbol $\sigma$ does not denote a generic
twist f\/ield, but rather the primary twist f\/ield as described
here.}, the f\/irst one being bosonic, the second fermionic. In
the sense of quantum chains, the f\/irst one is an ``order''
f\/ield, with non-zero vacuum expectation value, the second is a
``disorder'' f\/ield, with zero vacuum expectation value. As we
explained in sub-section \ref{semilocal}, to each of these
f\/ields there are two operators on~${\cal H}$, which makes four
operators: $\sigma_\pm$ and $\mu_\pm$. They are fully
characterised by the leading terms in the (equal-time) OPE's that
are displayed in Appendix \ref{app}. These leading terms are
f\/ixed by the general requirements (\ref{semilocp}) and
(\ref{semilocm}), by our choice of branch which says that when
fermion operators are placed before the twist-f\/ield operators,
they are on the same branch no matter the direction of the cut,
and by the general ``f\/ield'' product expansion that holds inside
correlation functions:
\begin{gather*}
    \psi(x,t-i0^+) \sigma(0,t) \sim \frc{i}{2\sqrt{\pi x+i0^+}} \mu(0,t) ,\qquad
    \psi(x,t-i0^+) \mu(0,t) \sim \frc{1}{2\sqrt{\pi x+i0^+}} \si(0,t),\\
    \b\psi(x,t-i0^+) \sigma(0,t) \sim -\frc{i}{2\sqrt{\pi x-i0^+}} \mu(0,t),\qquad
    \b\psi(x,t-i0^+) \mu(0,t) \sim \frc{1}{2\sqrt{\pi x-i0^+}} \si(0,t)
\end{gather*}
with branch cuts on $x<0$.

It is worth nothing that the relations of Appendix~\ref{app} are
in agreement with the Hermiticity relations $\si_\pm^\dag =
\si_\pm$ and $\mu_\pm^\dag =\pm \mu_\pm$.

\subsection[Riemann-Hilbert problem for twisted and untwisted
finite-temperature form factors]{Riemann--Hilbert problem for
twisted\\ and untwisted f\/inite-temperature form factors}

\subsubsection{Untwisted case}
\label{RHft}

In \cite{I}, the (untwisted) f\/inite-temperature form factors
(\ref{ftff}) of the twist-f\/ield operators above were shown to
solve a Riemann--Hilbert problem of the type found at zero
temperature, but with important modif\/ications. We repeat here
the results.

Consider the function
\[
    f_\eta(\theta_1,\ldots,\theta_k) = F^{\Or_\eta}_{+,\ldots,+}(\theta_1,\ldots,\theta_k;\beta)
\]
where $\Or_\eta$ is the operator with branch cut on its right
($\eta=+$) or on its left ($\eta=-$) representing any twist
f\/ield: this can be the order f\/ield $\sigma_\pm$ or the
disorder f\/ield $\mu_\pm$, or any of their conformal descendants
(that is, f\/ields which reproduce conformal descendants in the
massless limit). Conformal descendants include space derivatives,
as well as other f\/ields related to action of higher conformal
Virasoro modes on twist f\/ields. A way of describing such
descendants is by taking the limit $x\to0$ of the f\/inite part of
the OPE $\Or(x)\sigma_\pm(0)$ or $\Or(x)\mu_\pm(0)$, where $\Or$
is any bosonic operator formed out of normal-ordered products of
fermion operators.

The function $f$ solves the following Riemann--Hilbert problem:
\begin{enumerate}\itemsep=0pt
\item Statistics of free particles: $f$ acquires a sign under
exchange of any two of the rapidity variables; \item
Quasi-periodicity:
    \[
    f_\eta(\theta_1,\ldots,\theta_j+2i\pi,\ldots,\theta_k)
     = - f(\theta_1,\ldots,\theta_j,\ldots,\theta_k),\qquad j=1,\ldots,k;
    \]
\item Analytic structure: $f$ is analytic as function of all of
its variables $\theta_j,\,j=1,\ldots,k$ everywhere on the complex
plane except at simple poles. In the region ${\rm Im}(\theta_j)
\in [-i\pi,i\pi]$, $j=1,\ldots,k$, its analytic structure is
specif\/ied as follows:
\begin{enumerate}\itemsep=0pt
\item Thermal poles and zeroes: $f_\eta(\theta_1,\ldots,\theta_k)$
has poles at
    \[
    \theta_j=\alpha_n -\eta \frc{i\pi}2 ,\qquad n\in\Z,\quad j=1,\ldots,k
    \]
    and has zeroes at
    \[
    \theta_j = \alpha_n-\eta\frc{i\pi}2,\qquad n\in\Z+\frc12,\quad j=1,\ldots,k,
    \]
where $\alpha_n$ are def\/ined in (\ref{alphan}) (and, of course,
we use this def\/inition for any $n$); \item Kinematical
    poles: $f_\eta(\theta_1,\ldots,\theta_k)$ has poles, as a function of $\theta_k$, at
    $\theta_{j}\pm i\pi$, $j=1,\ldots,k-1$ with residues given by
    \[
    f_\eta(\theta_1,\ldots,\theta_k)
    \sim \pm\eta\frc{(-1)^{k-j}}\pi
    \frc{1+e^{-\beta E_{\theta_{j}}}}{1-e^{-\beta E_{\theta_{j}}}}
    \frc{f_\eta(\theta_1,\ldots,\h\theta_j,\ldots,\theta_{k-1})}{\theta_k-\theta_{j}\mp
    i\pi}.
    \]
\end{enumerate}
\end{enumerate}

In order to have other f\/inite-temperature form factors than
those with all positive charges, one more relation needs to be
used. We have:
\begin{itemize}\itemsep=0pt
\item[4.] Crossing symmetry:
\[
    F_{\ep_1,\ldots,\ep_j,\ldots,\ep_k}^{\Or_\eta}(\theta_1,\ldots,\theta_j+i\pi,\ldots,\theta_k;\beta)
    = i
    F_{\ep_1,\ldots,-\ep_j,\ldots,\ep_k}^{\Or_\eta}(\theta_1,\ldots,\theta_j,\ldots,\theta_k;\beta).
\]
\end{itemize}

The name ``crossing symmetry'' is inspired by the zero-temperature
case (and it is not to be confused with the simpler ``crossing
relations'' introduced in (\ref{crossingL}), (\ref{crossingR}),
(\ref{crossingLd}), (\ref{crossingRd})). To make it more obvious,
def\/ine the functions
\begin{gather}\label{ftme}
    f_\eta(\theta_1',\ldots,\theta_l'|\theta_1,\ldots,\theta_k) =
    (d_{+,\ldots,+}(\theta_1',\ldots,\theta_l'),
    \phi_L(\Or_\eta(0,0))\,d_{+,\ldots,+}(\theta_1,\ldots,\theta_k).
\end{gather}
These are in fact distributions, and can be decomposed in terms
supported at separated rapidities $\theta_i'\neq\theta_j$,
$\forall\,i,j$, and terms supported at colliding rapidities,
$\theta_i'=\theta_j$ for some $i$ and $j$. We will denote the
former by $f_\eta^{\rm
sep.}(\theta_1',\ldots,\theta_l'|\theta_1,\ldots,\theta_k)$, and
the latter by $f_\eta^{\rm
coll.}(\theta_1',\ldots,\theta_l'|\theta_1,\ldots,\theta_k)$.
Under integration over rapidity variables, the former gives
principal value integrals. Recalling the property
(\ref{crossingLd}), we have
\[
    f_\eta^{\rm sep.}(\theta_1',\ldots,\theta_l'\,|\,\theta_1,\ldots,\theta_k)
    =
    F^{\Or_\eta}_{+,\ldots,+,-,\ldots,-}(\theta_1,\ldots,\theta_k,\theta_l',\ldots,\theta_1';\beta)
\]
for $(\theta_i'\neq\theta_j$
$\forall\,i\in\{1,\ldots,l\},\,j\in\{1,\ldots,k\})$, where on the
right-hand side, there are $k$ positive charges $(+)$, and $l$
negative charges $(-)$. Analytically extending from its support
the distribution $f_\eta^{\rm sep.}$ to a function of complex
rapidity variables, crossing symmetry can then be written
\begin{gather*}
    f_\eta^{\rm sep.}(\theta_1',\ldots,\theta_l'|\theta_1,\ldots,\theta_k+i\pi)
    = i
    f_\eta^{\rm sep.}(\theta_1',\ldots,\theta_l',\theta_k|\theta_1,\ldots,\theta_{k-1}),
\\
    f_\eta^{\rm sep.}(\theta_1',\ldots,\theta_l'+i\pi|\theta_1,\ldots,\theta_k)
    = i
    f_\eta^{\rm sep.}(\theta_1',\ldots,\theta_{l-1}'|\theta_1,\ldots,\theta_k,\theta_l'),
\end{gather*}
which justif\/ies its name.

It is worth mentioning that the distributive terms corresponding
to colliding rapidities satisfy a set of recursive equations:
\begin{itemize}\itemsep=0pt
\item[5.] Colliding part of matrix elements:
\begin{gather*}
    f_\eta^{\rm coll.}(\theta_1',\ldots,\theta_l'|\theta_1,\ldots,\theta_k)\\
    \quad{}    =
    \sum_{i=1}^l\sum_{j=1}^k (-1)^{l+k-i-j}\frc{1+e^{-\beta E_{\theta_j}}}{1-e^{\beta E_{\theta_j}}}
    \,\delta(\theta_i'-\theta_j) f_\eta(\theta_1',\ldots,\h\theta_i',\ldots,\theta_l'|
    \theta_1,\ldots,\h\theta_j,\ldots,\theta_k).
\end{gather*}
\end{itemize}

Note that the colliding part vanishes in the limit of zero
temperature, $\beta\to\infty$. Finally, it is instructive to
re-write the distribution
$f_\eta(\theta_1',\ldots,\theta_l'\,|\,\theta_1,\ldots,\theta_k)$
as an analytical function with slightly shifted rapidities, plus a
distribution, using the relations
\begin{gather}\label{relppdelta}
    \frc1{\theta\mp i0^+} = \pm i\pi\delta(\theta) +
    \prin\lt(\frc1\theta\rt),
\end{gather}
where $\prin$ means that we must take the principal value integral
under integration. Def\/ining the disconnected part $f_\eta^{\rm
disconn.}(\theta_1',\ldots,\theta_l'\,|\,\theta_1,\ldots,\theta_k)$
of the matrix element (\ref{ftme}) as
\begin{gather*}
    f_\eta(\theta_1',\ldots,\theta_l'\,|\,\theta_1,\ldots,\theta_k)
    = f_\eta^{\rm sep.}(\theta_1'-\eta i0^+,\ldots,\theta_l'-\eta i0^+\,|\,\theta_1,\ldots,\theta_k)\\
    \phantom{f_\eta(\theta_1',\ldots,\theta_l'\,|\,\theta_1,\ldots,\theta_k)=}{}
    +
    f_\eta^{\rm disconn.}(\theta_1',\ldots,\theta_l'\, |\, \theta_1,\ldots,\theta_k),
\end{gather*}
where again we analytically extend from its support the
distribution $f_\eta^{\rm sep.}$ to a function of complex rapidity
variables, we f\/ind that the disconnected part satisf\/ies the
recursion relations
\begin{gather*}
    f_\eta^{\rm disconn.}(\theta_1',\ldots,\theta_l'\,|\,\theta_1,\ldots,\theta_k)
   \\
   \qquad{}=
    \sum_{i=1}^l\sum_{j=1}^k (-1)^{l+k-i-j}\,(1+e^{-\beta E_{\theta_j}})
    \,\delta(\theta_i'-\theta_j) f_\eta(\theta_1',\ldots,\h\theta_i',\ldots,\theta_l'\,|\,
    \theta_1,\ldots,\h\theta_j,\ldots,\theta_k).
\end{gather*}
Note that the factor $(1+e^{-\beta E_{\theta_j}})
\,\delta(\theta_i'-\theta_j)$ appearing inside the double sum is
just the overlap $(d_{+}(\theta_i'),d_{+}(\theta_j))$, so that the
equation above can be naturally represented as a ``sum of
disconnected diagrams.'' This equation is, in fact, consequence of
the general relation (\ref{crossingLdgen}).

\subsubsection{Twisted case}
\label{RHfttw}

The twisted case was not considered in \cite{I}, but can be
obtained from the same arguments.

There is no $U(1)$ invariance, but we can still twist by the
$\Z_2$ symmetry. Hence, we consider twisted f\/inite-temperature
form factors (\ref{twftff}) with $\om = \frc12$. The derivation of
\cite{I} for the Riemann--Hilbert problem can easily be adapted to
this case, and the results are as follows.

Consider the function
\[
    f_\eta(\theta_1,\ldots,\theta_k) = \le{\frc12}F^{\Or_\eta}_{+,\ldots,+}(\theta_1,\ldots,\theta_k;\beta),
\]
where $\Or_\eta$ is the operator with branch cut on its right
($\eta=+$) or on its left ($\eta=-$) representing a twist f\/ield.
The function $f$ solves the following Riemann--Hilbert problem:
\begin{enumerate}\itemsep=0pt
\item Statistics of free particles: $f$ acquires a sign under
exchange of any two of the rapidity variables; \item
Quasi-periodicity:
    \[
    f_\eta(\theta_1,\ldots,\theta_j+2i\pi,\ldots,\theta_k)
     = - f_\eta(\theta_1,\ldots,\theta_j,\ldots,\theta_k),\qquad j=1,\ldots,k;
    \]
\item Analytic structure: $f$ is analytic as function of all of
its variables $\theta_j$, $j=1,\ldots,k$ every\-where on the
complex plane except at simple poles. In the region ${\rm
Im}(\theta_j) \in [-i\pi,i\pi]$, $j=1,\ldots,k$, its analytic
structure is specif\/ied as follows:
\begin{enumerate}\itemsep=0pt
\item Thermal poles and zeroes: $f_\eta(\theta_1,\ldots,\theta_k)$
has poles at
    \[
    \theta_j=\alpha_n -\eta \frc{i\pi}2 ,\qquad n\in\Z+\frc12,\quad j=1,\ldots,k
    \]
    and has zeroes at
    \[
    \theta_j = \alpha_n -\eta \frc{i\pi}2,\qquad n\in\Z,\quad j=1,\ldots,k;
    \]
\item Kinematical poles: $f_\eta(\theta_1,\ldots,\theta_k)$ has
poles,
    as a function of $\theta_k$, at
    $\theta_{j}\pm i\pi$, $j=1,\ldots,k-1$ with residues given by \[
    f_\eta(\theta_1,\ldots,\theta_k;L)
    \sim \pm\eta \frc{(-1)^{k-j}}\pi
    \frc{1-e^{-\beta E_{\theta_{j}}}}{1+e^{-\beta E_{\theta_{j}}}}
    \frc{f_\eta(\theta_1,\ldots,\h\theta_j,\ldots,\theta_{k-1})}{\theta_k-\theta_{j}\mp
    i\pi}.
    \]
\end{enumerate}
\end{enumerate}

Again, in order to have other f\/inite-temperature form factors
than those with all positive charges, one more relation needs to
be used. We have:
\begin{itemize}
\item[4.] Crossing symmetry:
\[
    \le{\frc12} F_{\ep_1,\ldots,\ep_j,\ldots,\ep_k}^{\Or_\eta}(\theta_1,\ldots,\theta_j+i\pi,\ldots,\theta_k;\beta)
    = i
    \le{\frc12} F_{\ep_1,\ldots,-\ep_j,\ldots,\ep_k}^{\Or_\eta}(\theta_1,\ldots,\theta_j,\ldots,\theta_k;\beta).
\]
\end{itemize}

Moreover, matrix elements
\[
    f_\eta(\theta_1',\ldots,\theta_l'|\theta_1,\ldots,\theta_k) =
    (d_{+,\ldots,+}(\theta_1',\ldots,\theta_l'),
    \phi_L(\Or_\eta(0,0))\;d_{+,\ldots,+}(\theta_1,\ldots,\theta_k)_{\frc12}
\]
can again be decomposed in terms supported at separated rapidities
$\theta_i'\neq\theta_j$, $\forall\,i,j$ (which give~principal
value integrals under integration), and terms supported at
colliding rapidities, \mbox{$\theta_i'=\theta_j$} for some $i$ and
$j$, denoted respectively by $f_\eta^{\rm
sep.}(\theta_1',\ldots,\theta_l'\,|\,\theta_1,\ldots,\theta_k)$
and $f_\eta^{\rm coll.}(\theta_1',\ldots,\theta_l'\,| $
$\theta_1,\ldots,\theta_k)$. Recalling the property
(\ref{crossingLd}), we have
\[
    f_\eta^{\rm sep.}(\theta_1',\ldots,\theta_l'\,|\,\theta_1,\ldots,\theta_k)
    =
    \le{\frc12}F^{\Or_\eta}_{+,\ldots,+,-,\ldots,-}(\theta_1,\ldots,\theta_k,\theta_l',\ldots,\theta_1';\beta)
\]
for
$(\theta_i'\neq\theta_j\;\forall\;i\in\{1,\ldots,l\},\,j\in\{1,\ldots,k\})$,
where on the right-hand side, there are $k$ positive charges
$(+)$, and $l$ negative charges $(-)$. The distributive terms
corresponding to colliding rapidities satisfy again a set of
recursive equations, now modif\/ied by the twisting:
\begin{itemize}
\item[5.] Colliding part of matrix elements:
\begin{gather*}
    f_\eta^{\rm coll.}(\theta_1',\ldots,\theta_l'\,|\,\theta_1,\ldots,\theta_k)
   \\
   \quad{}=
    \sum_{i=1}^l\sum_{j=1}^k (-1)^{l+k-i-j}\frc{1-e^{-\beta E_{\theta_j}}}{1+e^{\beta E_{\theta_j}}}
    \,\delta(\theta_i'-\theta_j) f_\eta(\theta_1',\ldots,\h\theta_i',\ldots,\theta_l'\,|\,
    \theta_1,\ldots,\h\theta_j,\ldots,\theta_k).
\end{gather*}
\end{itemize}

Finally, we can again re-write the distribution
$f_\eta(\theta_1',\ldots,\theta_l'\,|\,\theta_1,\ldots,\theta_k)$
as an analytical function with slightly shifted rapidities, plus a
distribution, using the relations (\ref{relppdelta}). Def\/ining
the disconnected part $f_\eta^{\rm
disconn.}(\theta_1',\ldots,\theta_l'\,|\,\theta_1,\ldots,\theta_k)$
of the matrix element (\ref{ftme}) as
\begin{gather*}
    f_\eta(\theta_1',\ldots,\theta_l'\,|\,\theta_1,\ldots,\theta_k)
    = f_\eta^{\rm sep.}(\theta_1'-\eta i0^+,\ldots,\theta_l'-\eta i0^+\,|\,\theta_1,\ldots,\theta_k)\\
    \phantom{f_\eta(\theta_1',\ldots,\theta_l'\,|\,\theta_1,\ldots,\theta_k)=}{}
    +
    f_\eta^{\rm disconn.}(\theta_1',\ldots,\theta_l'\,|\,\theta_1,\ldots,\theta_k),
\end{gather*}
where again we analytically extend from its support the
distribution $f_\eta^{\rm sep.}$ to a function of complex rapidity
variables, we f\/ind that the disconnected part satisf\/ies the
recursion relations
\begin{gather*}
    f_\eta^{\rm disconn.}(\theta_1',\ldots,\theta_l'\,|\,\theta_1,\ldots,\theta_k)
   \\
   \quad{}=
    \sum_{i=1}^l\sum_{j=1}^k (-1)^{l+k-i-j}\,(1-e^{-\beta E_{\theta_j}})
    \,\delta(\theta_i'-\theta_j) f_\eta(\theta_1',\ldots,\h\theta_i',\ldots,\theta_l'\,|\,
    \theta_1,\ldots,\h\theta_j,\ldots,\theta_k).
\end{gather*}

\subsubsection[Other local fields]{Other local f\/ields}

It is worth noting that points 1, 2 and 4 are in fact also valid
for f\/ields that are local with respect to $\psi$ and $\b\psi$.
The analytic structure, point 3, for such f\/ields, is much
simpler: the f\/inite-temperature form factors are entire
functions of all rapidities. In fact, the f\/inite-temperature
form factors of $\psi$ and $\b\psi$ themselves are exactly equal
to their zero-temperature form factors, and for other f\/ields, a
phenomenon of mixing occurs, as described in \cite{I} and as can
be calculated using the techniques of Section \ref{formalstruct}.

\subsubsection[Differences with zero-temperature Riemann--Hilbert problems, and some explanations]{Dif\/ferences
with zero-temperature Riemann--Hilbert problems,\\ and some
explanations}

There are three main dif\/ferences between the Riemann--Hilbert
problems stated in this sub-section, and the Riemann--Hilbert
problems solved by zero-temperature form factors, reviewed in
Section~\ref{review}. First, there are, in the former, so-called
``thermal'' poles and zeroes. They are in fact consequences of the
semi-locality of the operators with respect to the fundamental
fermion f\/ields, and play the role of ``changing the sector'' of
the excited states when integrals are deformed to reproduce the
form factor expansion in the quantization on the circle. Indeed,
they displace the poles of the measure in order to reproduce the
right set of discrete momenta.

Second, the kinematical residue has an additional factor. This
factor, in fact, is closely related to the presence of the thermal
poles and zeroes.

Finally, there is a subtle but important dif\/ference: the
quasi-periodicity equation has a sign dif\/ference. Essentially,
the quasi-periodicity equation that we have at
f\/inite-temperature is exactly the one we would have at
zero-temperature with f\/ields that are {\em local} with respect
to the fermion f\/ield. This dif\/ference is again due to the
presence of the thermal poles and zeroes. More precisely, in the
limit of zero temperature, the f\/inite-temperature form factors
converge to the zero temperature one only in the strip ${\rm
Im}(\theta) \in ]-\pi/2,\pi/2[$. At the lines ${\rm Im}(\theta) =
\pm \pi/2$ (the sign depending on which excitation type and form
factor we are looking at), there is an accumulation of poles and
zeroes that produces a cut. The quasi-periodicity equation of zero
temperature comes from the analytical continuation through this
cut. Note that it is this analytical continuation that recovers
rotation invariance in Euclidean plane, an invariance which is
broken by the cylindrical geometry at f\/inite temperature.

We would like to mention, in relation to the breaking of Euclidean
rotation invariance, that yet crossing symmetry, point 4, is
valid. It is in fact a consequence of the fact that the
deformation of the contours, as explained in
Subsection~\ref{ftcircle}, should give residues at the poles of
the measure occurring in (\ref{ftffexp}). These residues come from
two contributions: the contribution of the displaced~$\theta$
contour associated to $\ep=+$, and that associated to~$\ep=-$.
That these two contributions should give a residue impose certain
conditions on the value of the f\/inite-temperature form factors:
they should correspond to contours in opposite direction and on
opposite sides of the same pole. From this and from knowing that
all f\/inite-temperature form factors of the fermion f\/ields
satisfy crossing symmetry, one concludes that crossing symmetry
holds for all local f\/ields.

\subsection[Results for twisted and untwisted finite-temperature form factors]{Results
for twisted and untwisted f\/inite-temperature form factors}

\label{sectftfftwist}

Again, we repeat here the results of \cite{I}, and generalise them
to the twisted case. Note that the method of computing
one-particle f\/inite-temperature form factors by solving the
Riemann--Hilbert problem with this asymptotic  is very similar to
the method used by Fonseca and Zamolodchikov \cite{FonsecaZ01} for
calculating form factors on the circle.

For the order and disorder operators, $\sigma_\pm$ and $\mu_\pm$
the solutions to the Riemann--Hilbert problems above are
completely f\/ixed (up to a normalization) by the asymptotic
 $\sim O(1)$ at $\theta\to\pm\infty$, imposed by the fact that they are
primary f\/ields of spin 0.

For the one-particle f\/inite-temperature form factor of the
disorder operator with a branch cut on its right, the solution is
\begin{gather}\label{1pftffmu+}
    F_\pm^{\mu_+}(\theta;\beta) = e^{\pm\frc{i\pi}4} C(\beta)\,
    \exp\lt[\mp\int_{-\infty\mp i0^+}^{\infty \mp i0^+} \frc{d\theta'}{2\pi i}
    \frc1{\sinh(\theta-\theta')}
    \ln\lt(\frc{1+e^{-\beta E_{\theta'}}}{1-e^{-\beta E_{\theta'}}}\rt)\rt]
\end{gather}
for some real constant $C(\beta)$. This is in agreement with the
Hermiticity of $\mu_+$, which gives
$(F_\pm^{\mu_+}(\theta;\beta))^* = F_\mp^{\mu_+}(\theta;\beta)$
for $\theta$ real. Using
\[
    \frc1{\sinh(\theta-(\theta'\pm i0^+))} = \pm i\pi \delta(\theta-\theta') +
    \prin\lt(\frc1{\sinh(\theta-\theta')}\rt)
\]
this can also be written
\begin{gather*}
    F_\pm^{\mu_+}(\theta;\beta)
    = C(\beta) e^{\pm\frc{i\pi}4}
    \sqrt{\frc{1+e^{-\beta E_\theta}}{1-e^{-\beta E_\theta}}}\;
    \exp\lt[\mp\int_{-\infty}^\infty \frc{d\theta'}{2\pi i}
    \prin\lt(\frc{1}{\sinh(\theta-\theta')}\rt)
    \ln\lt(\frc{1+e^{-\beta E_{\theta'}}}{1-e^{-\beta E_{\theta'}}}\rt)\rt].
\end{gather*}
That this is a solution can be checked by verifying the asymptotic
 $F_\pm^{\mu_+}(\theta;\beta) \sim e^{\pm\frc{i\pi}4}C(\beta)$ as
$|\theta|\to\infty$, and by verifying that the functions
$F_\pm^{\mu_+}(\theta;\beta)$ have poles and zeros at the proper
positions. Positions of poles and zeros are the values of $\theta$
such that when analytically continued from real values, a pole at
$\sinh(\theta-\theta')=0$ in the integrand of (\ref{1pftffmu+})
and one of the logarithmic branch points pinch the $\theta'$
contour of integration. The fact that these positions correspond
to poles and zeros can be deduced most easily from the functional
relation
\begin{gather}\label{fcteqn}
    F_\pm^{\mu_+}(\theta;\beta)F_\pm^{\mu_+}(\theta\pm i\pi;\beta) = \pm i C(\beta)^2
    \frc{1+e^{-\beta E_\theta}}{1-e^{-\beta E_\theta}}.
\end{gather}
Note that this implies the quasi-periodicity property
\[
    F_\pm^{\mu_+}(\theta+2i\pi;\beta) = -F_\pm^{\mu_+}(\theta;\beta).
\]
It is also easy to see that the crossing symmetry relation is
satisf\/ied.

For the operator $\mu_-$ with a branch cut on its left, one can
check similarly that the function
\[
    F_\pm^{\mu_-}(\theta;\beta) = F_\pm^{\mu_+}(\theta-i\pi;\beta) =
    -iF_\mp^{\mu_+}(\theta;\beta)
\]
solves the Riemann--Hilbert problem of Paragraph~\ref{RHft} with
$\eta=-$. Explicitly,
\begin{gather}\label{1pftffmu-}
    F_\pm^{\mu_-}(\theta;\beta) = -ie^{\mp\frc{i\pi}4} C(\beta)\,
    \exp\lt[\pm\int_{-\infty\pm i0^+}^{\infty \pm i0^+} \frc{d\theta'}{2\pi i}
    \frc1{\sinh(\theta-\theta')}
    \ln\lt(\frc{1+e^{-\beta E_{\theta'}}}{1-e^{-\beta E_{\theta'}}}\rt)\rt].
\end{gather}
In particular, we observe that $(F_\pm^{\mu_-}(\theta;\beta))^* =
-F_\mp^{\mu_-}(\theta;\beta)$, which is in agreement with the
anti-Hermiticity of the operator $\mu_-$. Note that we chose the
same constant $C(\beta)$ as a normalization for both
$F_\pm^{\mu_-}$ and $F_\pm^{\mu_+}$. This is not a consequence of
the Riemann--Hilbert problem, but can be checked by explicitly
calculating the normalisation. The normalisation was calculated in
\cite{I}, and is given by
\begin{gather}\label{normC}
    C(\beta) = \frc{\braL \sigma \ketL}{\sqrt{2\pi}},
\end{gather}
where the average $\braL\sigma\ketL$ was calculated in
\cite{Sachdev96} (the average at zero-temperature (that is,
$\beta\to\infty$) can be found in \cite{McCoyWu}) and is given by
\begin{gather*}
    m^{\frc18}2^{\frc1{12}} e^{-\frc18} A^{\frc32} \exp\lt[ \frc{(m\beta)^2}2
    \int\int_{-\infty}^\infty \frc{d\theta_1d\theta_2}{(2\pi)^2}
    \frc{\sinh\theta_1\sinh\theta_2}{\sinh(m\beta\cosh\theta_1)\sinh(m\beta\cosh\theta_2)}\right.\\
\left.\qquad{}\times
    \ln\lt|\lt(\coth\frc{\theta_1-\theta_2}2\rt)\rt|\rt],
\end{gather*}
where $A$ is Glaisher's constant.
 Essentially, this normalisation is evaluated by computing
the~lea\-ding  of $\braL \psi(x,0) \mu_-(0,0)\ketL$ as $x\to0^+$,
and the leading
 of $\braL \mu_+(0,0) \psi(x,0) \ketL$ as $x\to0^-$, using the form factor
expansions; in both cases, it is important to approach the point
$x=0$ from a region that is away from the cut.

Multi-particle f\/inite-temperature form factors can be easily
constructed from the well-known zero-temperature form factors
(f\/irst calculated in \cite{BergKW79}), by adjoining ``leg
factors'', which are just normalized one-particle
f\/inite-temperature form factors:
\[
    F^{\Or_+}_{+,\ldots,+}(\theta_1,\ldots,\theta_k;\beta)
    = i^{\lt[\frc{k}2\rt]}  \braL\sigma\ketL
    \lt(\prod_{j=1}^k
    \frc{F^{\mu_+}_+(\theta_j;\beta)}{\braL\sigma\ketL}
    \rt)
     \prod_{1\leq i<j\leq k}
    \tanh\lt(\frc{\theta_j-\theta_i}2\rt),
\]
where $\Or_+$ is $\sigma_+$ if $k$ is even, and $\mu_+$ if $k$ is
odd. The symbol $[k/2]$ equals the greatest integer smaller than
or equal to $k/2$. This satisf\/ies the condition on thermal poles
and zeroes simply from the properties of the leg factors, and it
can be verif\/ied that this satisf\/ies the quasi-periodicity
condition and the kinematical pole condition, Point 2 and Point 3b
of Subsection~\ref{RHft}, respectively. Using crossing symmetry,
Point 4, it is a simple matter to obtain the formula for other
values of the charges:
\begin{gather}\label{ftfftwistp}
    F^{\Or_+}_{\ep_1,\ldots,\ep_k}(\theta_1,\ldots,\theta_k;\beta)
    = i^{\lt[\frc{k}2\rt]}  \braL\sigma\ketL
    \lt(\prod_{j=1}^k
    \frc{F^{\mu_+}_{\ep_j}(\theta_j;\beta)}{\braL\sigma\ketL}
    \rt)
    \prod_{1\leq i<j\leq k}
    \lt(\tanh\lt(\frc{\theta_j-\theta_i}2\rt)\rt)^{\ep_i\ep_j}\!\!.\!\!\!
\end{gather}

Similarly, we have
\begin{gather}\label{ftfftwistm}
    F^{\Or_-}_{\ep_1,\ldots,\ep_k}(\theta_1,\ldots,\theta_k;\beta)
    = i^{\lt[\frc{k}2\rt]}  \braL\sigma\ketL
    \lt(\prod_{j=1}^k
    \frc{F^{\mu_-}_{\ep_j}(\theta_j;\beta)}{\braL\sigma\ketL}\rt)
     \prod_{1\leq i<j\leq k}
    \lt(\tanh\lt(\frc{\theta_j-\theta_i}2\rt)\rt)^{\ep_i\ep_j}\!\!,\!\!\!
\end{gather}
where $\Or_-$ is $\sigma_-$ if $k$ is even, and $\mu_-$ if $k$ is
odd.

Finally, twisted one-particle f\/inite-temperature form factors
can easily be obtained by solving the Riemann--Hilbert problem of
Paragraph~\ref{RHfttw} as follows:
\[
    \le{\frc12}F_\ep^{\mu_\eta}(\theta) = \frc{\ep i C(\beta)^2}{F_\ep^{\mu_\eta}(\theta)}.
\]
These functions have the correct analytic structure, they satisfy
the crossing symmetry relation (point 4), and their normalisation
is the correct one that can be deduced from the fact that the
leading  of $( \psi(x,0) ,\mu_-(0,0))_{\frc12}$ as $x\to0^+$, and
the leading
 of $( \mu_+(0,0), \psi(x,0) )_{\frc12}$ as $x\to0^-$, are the same as in the
untwisted case. Twisted multi-particle form factors can also be
obtained in a simple way:
\begin{gather}
    \le{\frc12}F^{\Or_\eta}_{\ep_1,\ldots,\ep_k}(\theta_1,\ldots,\theta_k;\beta)\nonumber\\
    \qquad{}    = i^{\lt[\frc{k}2\rt]}  \braL\sigma\ketL
    \lt(\prod_{j=1}^k
    \frc{\le{\frc12}F^{\mu_\eta}_{\ep_j}(\theta_j;\beta)}{\braL\sigma\ketL}
    \rt)
     \prod_{1\leq i<j\leq k}
    \lt(\tanh\lt(\frc{\theta_j-\theta_i}2\rt)\rt)^{\ep_i\ep_j}\!\!,\label{twftfftwist}
\end{gather}
where $\Or_\pm$ is $\sigma_\pm$ if $k$ is even, and $\mu_\pm$ if
$k$ is odd.

\subsection[Form factors on the circle from finite-temperature form factors]{Form
factors on the circle from f\/inite-temperature form factors}

As explained in Subsection \ref{ftcircle}, there is a relation
between f\/inite-temperature form factors and form factors in the
quantization on the circle. In the present case of the Majorana
theory, this relation was written explicitly in \cite{I}, and was
proven by independent means. A slight extension to the twisted
case gives it as follows:
\begin{gather}
    {}_\beta\bra\t{n}_1,\ldots,\t{n}_l|\h\Or(0,0)|n_1,\ldots,n_k\ket_\beta \nonumber\\
    \qquad{}= e^{-\frc{i\pi s}2}
        \lt(\frc{2\pi}{m L}\rt)^{\frc{k+l}2}
        \lt(\prod_{j=1}^l \frc1{\sqrt{\cosh(\alpha_{\t{n}_j})}}\rt)
        \lt(\prod_{j=1}^k
        \frc1{\sqrt{\cosh(\alpha_{n_j})}}\rt)\nonumber\\
        \qquad\phantom{{}={}}\times
        \le\om F^\Or_{+,\ldots,+,-,\ldots,-}\lt(\alpha_{n_1}+\frc{i\pi}2,\ldots,
        \alpha_{n_k}+\frc{i\pi}2,
        \alpha_{\t{n}_l}+\frc{i\pi}2,\ldots,\alpha_{\t{n}_1}+\frc{i\pi}2;\beta\rt),\label{relftcyl}
    \end{gather}
where there are $k$ positive charges and $l$ negative charges in
the indices of $\le\om F^\Or$, and where $\alpha_n$ are def\/ined
in (\ref{alphan}). Here, $s$ is the spin of $\Or$. This formula is
valid for any excited states in the sector above
$|\vac_{\frc12+\om}\ket$ (see the discussion around
(\ref{DeltaE})). That is, if $\om=0$, it is valid for excited
states in the NS vacuum, hence with $n_i,\t{n}_i\in \Z+\frc12$.
For $\om=\frc12$, it is valid for excited states in the R vacuum,
hence with $n_i,\t{n}_i \in \Z$.

When $\Or$ is a twist f\/ield, its associated branch cut changes
the sector of the bra or the ket, hence formula (\ref{relftcyl})
can then be applied only if one of the bra or the ket is the
vacuum, and if the branch cut associated to the twist f\/ield is
chosen so that this vacuum is in the opposite sector (in order to
keep the excited states in the same sector). If $\om=0$, the
vacuum will then be in the $R$ sector, and if $\om=\frc12$, it
will be in the $NS$ sector. For a branch cut to the right, it is
the bra that must be chosen as this vacuum, whereas for a branch
cut to the left, it is the ket.

It is easy to check, using (\ref{relftcyl}), that the formulas
above for f\/inite-temperature form factors reproduce the known
form factors on the circle \cite{Bugrij00,Bugrij01,FonsecaZ01}.

\subsection{Two-point functions, Fredholm determinants\\ and scaling limit of the quantum Ising model}

\label{twoptfct}

The f\/inite-temperature form factor expansion (\ref{twftffexp})
now gives explicit expansions for f\/inite-temperature two-point
functions of twist f\/ields at $x>0$:
\begin{gather}
    \braL\sigma_+(x,t)\sigma_-(0,0)\ketL^\om = e^{\lt(\Evac[\om] - \Evac\lt[\frc12+\om\rt]\rt)x}\nonumber\\
    \qquad{}    \times \sum_{k=0\atop k\ {\rm even}}^\infty \sum_{\ep_1,\ldots,\ep_k=\pm}
    \int_{\{{\rm Im}(\theta_j)=\ep_j 0^+\}} \frc{d\theta_1\cdots
    d\theta_k e^{\sum\limits _{j=1}^k\ep_j (imx\sinh\theta_j-imt\cosh\theta_j)}}{
    k! \prod\limits _{j=1}^k\lt(1+ e^{2\pi i \om} e^{-\ep_j
    m\beta\cosh\theta_j}\rt)}\nonumber\\
    \qquad{}\times     i^k \prod_{j=1}^k (\le\om F^{\mu_+}_{\ep_j}(\theta_j;L))^2
    \prod_{1\leq i<j\leq k}
    \tanh\lt(\frc{\theta_j-\theta_i}2\rt)^{2\ep_i\ep_j}\label{R1}
    \end{gather}
and
\begin{gather}
    \braL\mu_+(x,t)\mu_-(0,0)\ketL^\om = - e^{\lt(\Evac[\om] - \Evac\lt[\frc12+\om\rt]\rt)x}\nonumber\\
\qquad{} \times \sum_{k=0\atop k\ {\rm odd}}^\infty
\sum_{\ep_1,\ldots,\ep_k=\pm}
    \int_{\{{\rm Im}(\theta_j)=\ep_j 0^+\}} \frc{d\theta_1\cdots
    d\theta_k e^{\sum\limits_{j=1}^k\ep_j (imx\sinh\theta_j-imt\cosh\theta_j)}}{
    k!\prod\limits_{j=1}^k\lt(1+ e^{2\pi i \om} e^{-\ep_j
    m\beta\cosh\theta_j}\rt)}\nonumber\\
    \qquad{}\times
    i^k \prod_{j=1}^k (\le\om F^{\mu_+}_{\ep_j}(\theta_j;L))^2
    \prod_{1\leq i<j\leq k}
    \tanh\lt(\frc{\theta_j-\theta_i}2\rt)^{2\ep_i\ep_j},\label{R2}
\end{gather}
where $\om=0$ or $\om=\frc12$, and we recall that $\Evac[0] =
\Evac_{{\rm R}}$ and $\Evac[1/2] = \Evac_{{\rm NS}}$ are given in
(\ref{vacener}). In order to fully clarify the meaning of these
f\/inite-temperature correlation functions, we recall also that at
imaginary time $t=i\rx$ and at positive $x=\tau$, they correspond
to the following correlation functions in the quantization on the
circle:
\begin{gather*}
    \braL \sigma_+(\tau,i\rx) \sigma_-(0,0) \ketL^\om = {\ }_{\beta}\bra \vac_{\om}| \sigma(\rx,\tau)
    \sigma(0,0) |\vac_{\om}\ket_{\beta},\\
    \braL \mu_+(\tau,i\rx) \mu_-(0,0) \ketL^\om = {\ }_{\beta}\bra \vac_{\om}| \mu(\rx,\tau)
    \mu(0,0) |\vac_{\om}\ket_{\beta},
\end{gather*}
where the vacuum is in the R sector if $\om=0$, and in the NS
sector if $\om=1/2$.

Following \cite{I}, where techniques from
\cite{BabelonB92,LeclairLSS96} were borrowed, Fredholm determinant
representations can now easily be obtained for two-point functions
from the formulas
\begin{gather}\label{f1}
    \det_{i,j} \lt\{ \frc{u_i-u_j}{u_i+u_j} \rt\} =
    \lt\{\ba{ll} \prod_{1\leq i<j\leq k}
    \lt(\frc{u_i-u_j}{u_i+u_j}\rt)^2
    & k \ {\rm even}, \\
    0 & k \ {\rm odd}
     \ea
    \rt.
\end{gather}
and
\begin{gather}\label{f2}
    {\rm det}_{i,j} \lt\{\frc1{u_i+u_j}\rt\} =
    \frc1{2^ku_1\cdots u_k} \prod_{1\leq i<j\leq k}
    \lt(\frc{u_i-u_j}{u_i+u_j}\rt)^2.
\end{gather}
Formula (\ref{f1}) gives
\[
    \braL\sigma_+(x,t)\sigma_-(0,0)\ketL^\om
    = {\rm det}({\bf 1} + {\bf K}),
\]
where ${\bf K}$ is an integral operator with an additional index
structure, def\/ined by its action $({\bf K} f)_\ep(\theta) =
\sum\limits_{\ep'=\pm} \int_{-\infty}^{\infty} d\theta'
K_{\ep,\ep'}(\theta,\theta')f_{\ep'}(\theta')$ and its kernel
\[
    K_{\ep,\ep'}(\theta,\theta') =
    i (\le\om F_\ep^{\mu_+}(\theta;\beta))^2
    \tanh\lt(\frc{\theta'-\theta}2\rt)^{\ep\ep'}
    \frc{e^{\ep(imx\sinh\theta-imt\cosh\theta)}}{1+e^{2\pi i \om} e^{-\ep
    m\beta\cosh\theta}}.
\]
Finally, in order to obtain two-point functions of disorder
f\/ields, we must consider the linear combinations $\sigma\pm
\mu$. Formula (\ref{f2}) gives
\[
    \braL(\sigma_+(x,t)+\eta\mu_+(x,t))(\sigma_-(0,0)+\eta\mu_-(0,0))\ketL^\om
    = {\rm det}({\bf 1} + {\bf J}^{(\eta)})
\]
with $\eta=\pm$ and by def\/inition $({\bf J}^{(\eta)}f)_\ep(u) =
\sum\limits_{\ep'=\pm} \int_{0}^{\infty} du'
J^{(\eta)}_{\ep,\ep'}(u,u')f_{\ep'}(u')$ where the kernel is given
by
\[
    J^{(\eta)}_{\ep,\ep'}(u,u') =
    -2\eta\, i (\le\om F_\ep^{\mu_+}(\ln(u);\beta))^2
    \frc1{\ep u + \ep' u'}
    \frc{e^{\frc{\ep}2(i m x(u-u^{-1})-im t (u+u^{-1}))}}{1+e^{2\pi i \om} e^{-\frc{\ep
    m \beta}2(u+u^{-1})}}.
\]
The interest in Fredholm determinant representations is, in part,
that they can be used to ef\/f\/iciently obtain asymptotics of
correlation functions.

Finally, we mention that these two-point functions in the Majorana
theory can be used to evaluate the of\/f-critical scaling limit of
two-point functions in the quantum Ising chain (see, for instance,
the book \cite{ItzyksonDrouffe}). The quantum Ising chain is a
quantum mechanical model with Hamiltonian
\[
    H_{{\rm Ising}} = -\sum_j (J s^z_j s^z_{j+1} + h s_j^x)
\]
with $J>0$. The spin variables $s^x_j$ and $s^z_j$ are in the
spin-$1/2$ representation of $SU(2)$, and are two of the usual
Pauli matrices on the $j^{\rm th}$ two-dimensional space, the
third one being $s^y_j$:
\[
    s^x = \mato{cc} 0 & 1 \\ 1 & 0 \matf,\qquad s^y = \mato{cc} 0 & -i \\ i & 0 \matf,\qquad s^z=
    \mato{cc} 1 & 0 \\ 0 & -1 \matf.
\]
It is the ``Hamiltonian limit'' of the two-dimensional Ising
classical statistical model. There is a value $h=h_c$ of the
transverse magnetic f\/ield at which this model is critical. The
conformal f\/ield theory that describes it is the free massless
Majorana theory. For $h<h_c$, the system is ordered, and at zero
temperature the average of $s^z_j$ is non-zero. On the other hand,
for $h>h_c$, the system is disordered. As $h$ is made to approach
$h_c$, the correlation length $\xi$ associated to the two-point
function $\bra s^z_js^z_0\ket$ diverges. The scaling limit is
obtained by looking at the situation where $h\to h_c$, while the
inverse temperature is made to diverge as $J\beta\propto \xi$, and
the distances between points in correlation functions are made to
diverge as $|j|\propto \xi$. The quantum f\/ield theory model that
describes the appropriately normalised correlation functions
obtained in this limit is the free massive Majorana theory, the
product of mass times position being equal to $mx = |j|/\xi$. If
$h$ is sent to $h_c$ from below (ordered regime), then we have the
correspondence
\[
    Z^{-1} \xi^{\frc14}
      \frc{\Tr\lt(e^{-\beta H_{{\rm Ising}}} s^z_j(t) s^z_0(0)\rt)}{\Tr\lt( e^{-\beta H_{{\rm Ising}}} \rt)} \to
      m^{-\frc14} \braL \sigma_+(x,t)\sigma_-(0,0) \ketL^{\frc12},
\]
where
\[
    s^{z}_j(t) = e^{-itH_{{\rm Ising}}} s^{z}_j e^{itH_{{\rm Ising}}}
\]
and $Z$ is a non-zero, non-universal number. On the other hand, if
$h$ is sent to $h_c$ from above (disordered regime), then
\[
    Z^{-1} \xi^{\frc14}
    \frc{\Tr\lt(e^{-\beta H_{{\rm Ising}}} s^z_j(t) s^z_0(0)\rt)}{\Tr\lt( e^{-\beta H_{{\rm Ising}}} \rt)} \to
      m^{-\frc14} \braL \mu_+(x,t)\mu_-(0,0) \ketL^{\frc12}.
\]
It is important to realise that the spin variables $s^{z}$ does
not converge, in the scaling limit, to the twist f\/ields
$\sigma$, $\mu$; indeed, only its products converge to products of
twist f\/ields. This is clear, since the f\/inite-temperature
average of single twist f\/ields are non-zero (but have
non-trivial space dependence, as explained in
Subsection~\ref{semilocal}), but f\/inite-temperature averages of
spin variables are zero (since at f\/inite temperature, there can
be no symmetry breaking). One should recall that the passage from
the quantum Ising model to the Majorana theory involves writing
the spin variables as exponentials of sums of (bilinear of)
fermionic variables lying on a segment of the chain, and the two
end-points of the segment correspond to two spin variables.

\section{Perspectives}

We have developed partly the concept of f\/inite-temperature form
factor in the general context of factorised scattering theory, and
we completed the program in the case of the Majorana theory. The
most important next step is, of course, to complete this program
in models with non-trivial scattering. We believe that ideas
concerning the relation between f\/inite-temperature form factors
and matrix elements in the quantization on the circle will lead to
restrictive conditions that will greatly help fully f\/ix
f\/inite-temperature form factors in interacting models. Also, the
operator implementing the generalisation of CFT's ``mapping to the
cylinder'' may be useful, and this method is not far from the
explicit construction of ``boundary-creating operator'' in
integrable boundary QFT. The generalisation to interacting models
is a very important step, and will open the way to results about
large-distance and large-time behaviours of correlation functions in
interacting, integrable models.

Another interesting avenue is to generalise the program to the
free Dirac theory; this should not pose any dif\/f\/iculties, and
will clarify the structure of f\/inite-temperature form factors of
more general twist f\/ields (two-point functions at
f\/inite-temperature were already studied in~\cite{Lisovyy05}).
Then, it would be interesting to understand the structure for
descendants of twist fields in such free fermionic models,
perhaps using the operator $\mft$ def\/ined in (\ref{condmft})
that provides a~``mapping to the cylinder''.

Finally, one would like to obtain the full large-time expansion of
correlation functions in the quantum Ising model. Besides directly
using the f\/inite-temperature form factor expansion, it is
possible that the Fredholm determinant representations obtained
here can be used fruitfully for this purpose (work is in progress
\cite{DoyonGamsa06}).

\appendix

\section{OPE's in the Majorana theory}
\label{app}

The order and disorder f\/ields have operator representations
$\sigma_\pm$ and $\mu_\pm$ on ${\cal H}$. These operators are
completely characterised by the leading terms in their OPEs with
the fermion f\/ields:
\begin{alignat*}{3}
    &\psi(x,t) \sigma_+(0,t) \sim \frc{1}{2\sqrt{-\pi x-i0^+}} \mu_+(0,t) ,\qquad&&
    \sigma_+(0,t) \psi(x,t) \sim \frc{1}{2\sqrt{-\pi x+i0^+}} \mu_+(0,t),&\\
    &\psi(x,t) \sigma_-(0,t) \sim \frc{i}{2\sqrt{\pi x+i0^+}} \mu_-(0,t) ,\qquad&&
    \sigma_-(0,t) \psi(x,t) \sim \frc{i}{2\sqrt{\pi x-i0^+}} \mu_-(0,t),&\\
    &\psi(x,t) \mu_+(0,t) \sim \frc{-i}{2\sqrt{-\pi x-i0^+}} \si_+(0,t) ,\qquad&&
    \mu_+(0,t) \psi(x,t) \sim \frc{i}{2\sqrt{-\pi x+i0^+}} \si_+(0,t),&\\
    &\psi(x,t) \mu_-(0,t) \sim \frc{1}{2\sqrt{\pi x+i0^+}} \si_-(0,t),\qquad&&
    \mu_-(0,t) \psi(x,t) \sim \frc{-1}{2\sqrt{\pi x-i0^+}} \si_-(0,t)&
\end{alignat*}
and
\begin{alignat*}{3}
   & \b\psi(x,t) \sigma_+(0,t) \sim \frc{1}{2\sqrt{-\pi x+i0^+}} \mu_+(0,t) ,\qquad&&
    \sigma_+(0,t) \b\psi(x,t) \sim \frc{1}{2\sqrt{-\pi x-i0^+}} \mu_+(0,t) ,&\\
   & \b\psi(x,t) \sigma_-(0,t) \sim -\frc{i}{2\sqrt{\pi x-i0^+}} \mu_-(0,t),\qquad &&
    \sigma_-(0,t) \b\psi(x,t) \sim -\frc{i}{2\sqrt{\pi x+i0^+}} \mu_-(0,t), &\\
    &\b\psi(x,t) \mu_+(0,t) \sim \frc{i}{2\sqrt{-\pi x+i0^+}} \si_+(0,t) ,\qquad &&
    \mu_+(0,t) \b\psi(x,t) \sim \frc{-i}{2\sqrt{-\pi x-i0^+}} \si_+(0,t), &\\
   & \b\psi(x,t) \mu_-(0,t) \sim \frc{1}{2\sqrt{\pi x-i0^+}} \si_-(0,t) ,\qquad &&
    \mu_-(0,t) \b\psi(x,t) \sim \frc{-1}{2\sqrt{\pi x+i0^+}} \si_-(0,t), &
\end{alignat*}
where everywhere, the square root is on its principal branch.

\subsection*{Acknowledgments}

I am grateful to F.~Essler for many useful discussions and
continued interest in this work, and to A.~Gamsa for reading
through the manuscript. I would like to acknowledge support from
an EPSRC (UK) post-doctoral fellowship (grant GR/S91086/01).

\pdfbookmark[1]{References}{ref}
\LastPageEnding


\begin{thebibliography}{99} 

\footnotesize\itemsep=0pt

\bibitem{Kapusta}
Kapusta J.I., Finite temperature f\/ield theory, 
Cambridge University Press, Cambridge, 1989.

\bibitem{I}
Doyon B., Finite-temperature form factors in the free Majorana theory,
{\it J. Stat. Mech. Theory Exp.} (2005), P11006, 45 pages,
\href{http://arxiv.org/abs/hep-th/0506105}{\mbox{hep-th/0506105}}.

\bibitem{BourbonnaisJ99}
Bourbonnais C., Jerome D., The normal phase of
quasi-one-dimensional organic superconductors, in Advances in
Synthetic Metals, Twenty Years of Progress in Science and
Technology, Editors P.~Bernier, S.~Lefrant and E.~Bidan, Elsevier, New York,
 1999.

\bibitem{Gruner}
 Gruner G., Density waves in solids, Addison-Wesley,  Reading (MA), 1994.

\bibitem{EsslerK04}
Essler F.H.L., Konik R.M., Applications of massive integrable
quantum f\/ield theories to problems in condensed matter physics,
in From Fields to Strings: Circumnavigating Theoretical Physics,
Editors M.~Shifman, A.~Vainshtein and J.~Wheater, {\it Ian Kogan
Memorial Collection}, World Scientif\/ic, 2004.

\bibitem{VergelesG76} Vergeles S.N., Gryanik V.M.,
Two-dimensional quantum f\/ield theories having exact solutions,
{\it Yad. Fiz.} {\bf 23} (1976), 1324--1334 (in Russian).

\bibitem{Weisz77}
 Weisz P., Exact quantum sine-Gordon soliton form factors, {\it
Phys. Lett.~B} {\bf 67} (1977), 179--182.

\bibitem{KarowskiW78}
Karowski M., Weisz P., Exact form factors in (1+1)-dimensional
f\/ield theoretic models with soliton behaviour, {\it Nuclear
Phys. B} {\bf 139} (1978), 455--476.

\bibitem{BergKW79}
Berg B.,  Karowski M., Weisz P.,  Construction of Green's
functions from an exact $S$-matrix, {\it Phys. Rev. D} {\bf 19} (1979), 2477--2479.

\bibitem{Smirnov}
Smirnov F.A., Form factors in completely integrable models of
quantum f\/ield theory, World Scientif\/ic, Singapore, 1992.

\bibitem{ZamolodchikovAl91}
Zamolodchikov Al.B., Two-point correlation function in scaling
Lee--Yang model, {\it Nuclear Phys.~B} {\bf 348} (1991), 619--641.

\bibitem{Matsubara55}
Matsubara T.M., A new approach to quantum-statistical mechanics,
{\it Progr. Theoret. Phys.} {\bf 14} (1955), 351--378.


\bibitem{Kubo57}
Kubo R., Statistical-mechanical theory of irreversible processes.
I. General theory and simple applications to magnetic and
conduction problems, {\it J. Phys. Soc. Japan} {\bf 12} (1957),
570--586.

\bibitem{MartinS59}
Martin C., Schwinger J., Theory of many-particle systems. I, {\it
Phys. Rev.} {\bf 115} (1959), 1342--1373.

\bibitem{Smirnov98a}
Smirnov F.A., Quasi-classical study of form factors in f\/inite
volume,
\href{http://arxiv.org/abs/hep-th/9802132}{hep-th/9802132}.

\bibitem{Smirnov98b}
Smirnov F.A., Structure  of matrix elements in quantum Toda chain,
\href{http://arxiv.org/abs/hep-th/9805011}{hep-th/9805011}.

\bibitem{ElburgS00}
van Elburg R.A.J., Schoutens K., Form factors for quasi-particles
in $c=1$ conformal f\/ield theory, {\it J. Phys.~A: Math. Gen.} {\bf 33} (2000), 7987--8012,
\href{http://arxiv.org/abs/cond-mat/0007226}{cond-mat/0007226}.

\bibitem{MussardoRS03}
Mussardo G., Riva V., Sotkov G., Finite-volume form factors in
semiclassical approximation, {\it Nuclear Phys.~B} {\bf 670} (2003),
464--478,
\href{http://arxiv.org/abs/hep-th/0307125}{hep-th/0307125}.

\bibitem{Bugrij00}
Bugrij A.I., The correlation function in two dimensional Ising
model on the f\/inite size lattice.~I,
\href{http://arxiv.org/abs/hep-th/0011104}{\mbox{hep-th/0011104}}.

\bibitem{Bugrij01}
Bugrij A.I., Form factor representation of the correlation
function of the two dimensional Ising model on a~cylinder,
\href{http://arxiv.org/abs/hep-th/0107117}{hep-th/0107117}.

\bibitem{FonsecaZ01}
Fonseca P., Zamolodchikov A.B., Ising f\/ield theory in a magnetic
f\/ield: analytic properties of the free energy, {\it J. Statist.
Phys.} {\bf 110} (2003), 527--590,
\href{http://arxiv.org/abs/hep-th/0112167}{hep-th/0112167}.

\bibitem{LeplaeUM74}
Leplae L., Umezawa H., Mancini F., Derivation and application of
the boson method in superconductivity, {\it Phys. Rep.} {\bf 10} (1974), 151--272.

\bibitem{ArimitsuU87}
Arimitsu T., Umezawa H., Non-equilibrium thermo f\/ield dynamics,
{\it Prog. Theoret. Phys.} {\bf 77} (1987), 32--52.

\bibitem{ArimitsuU87a}
Arimitsu T., Umezawa H.,
 General structure of non-equilibrium thermo f\/ield dynamics, {\it Progr. Theoret. Phys.}
 {\bf 77} (1987), 53--67.

\bibitem{Henning95}
Henning P.A., Thermo f\/ield dynamics for quantum f\/ields with
continuous mass spectrum, {\it Phys. Rep.} {\bf 253} (1995), 235--381,
\href{http://arxiv.org/abs/nucl-th/9311001}{nucl-th/9311001}.

\bibitem{AmaralB05}
Amaral R.L.P.G., Belvedere L.V., Two-dimensional thermof\/ield
bosonization,
\href{http://arxiv.org/abs/hep-th/0504012}{hep-th/0504012}.

\bibitem{AltshulerT05}
Altshuler B.L., Konik R., Tsvelik A.M., Low temperature
correlation functions in integrable models: derivation of the
large distance and time asymptotics from the form factor
expansion, {\it Nuclear Phys. B} {\bf 739} (2006), 311--327,
\href{http://arxiv.org/abs/cond-mat/0508618}{cond-mat/0508618}.

\bibitem{Sachdev96}
Sachdev S., The universal, f\/inite temperature, crossover
functions of the quantum transition in the Ising chain in a
transverse f\/ield, {\it Nuclear Phys. B} {\bf 464} (1996), 576--595,
\href{http://arxiv.org/abs/cond-mat/9509147}{cond-mat/9509147}.

\bibitem{SachdevY97}
Sachdev S., Young A.P., Low temperature relaxational dynamics of
the Ising chain in a transverse f\/ield, {\it Phys. Rev. Lett.}
{\bf 78} (1997), 2220--2223,
\href{http://arxiv.org/abs/cond-mat/9609185}{cond-mat/9609185}.

\bibitem{KorepinBogoliubovIzergin93}
Korepin V.E., Bogoliubov N.M., Izergin A.G., Quantum inverse
scattering method and correlation functions, Cambridge
University Press, Cambridge, 1993.

\bibitem{DoyonGamsa06}
Doyon B., Gamsa A., Work  in progress.

\bibitem{Balog94}
Balog J., Field theoretical derivation of the TBA integral
equations, {\it Nuclear Phys.~B} {\bf 419} (1994), 480--512.

\bibitem{LeclairM99}
Leclair A., Mussardo G., Finite temperature correlation functions
in integrable QFT, {\it Nuclear Phys. B} {\bf 552} (1999), 624--642,
\href{http://arxiv.org/abs/hep-th/9902075}{hep-th/9902075}.

\bibitem{Lukyanov01}
Lukyanov S., Finite-temperature expectation values of local
f\/ields in the sinh-Gordon model, {\it Nuclear Phys.~B} {\bf 612} (2001), 391--412,
\href{http://arxiv.org/abs/hep-th/0005027}{hep-th/0005027}.

\bibitem{EsslerKComm} Essler F., Konik R., Private communication.

\bibitem{WuMTB76}
Wu T.T., McCoy B.M., Tracy C.A., Barouch E., Spin-spin correlation
functions for the two-dimensional Ising model: exact theory in the
scaling region, {\it Phys. Rev. B} {\bf 13} (1976), 316--374.

\bibitem{Perk80}
Perk J.H.H., Equations of motion for the transverse correlations
of the one-dimensional $XY$-model at f\/inite temperature, {\it
Phys. Lett. A} {\bf 79} (1980), 1--2.

\bibitem{Lisovyy02}
Lisovyy O., Nonlinear dif\/ferential equations for the correlation
functions of the 2D Ising model on the cylinder, {\it Adv. Theor.
Math. Phys.} {\bf 5} (2002), 909--922.

\bibitem{FonsecaZ03}
Fonseca P., Zamolodchikov A.B., Ward identities and integrable
dif\/ferential equations in the Ising f\/ield theory,
\href{http://arxiv.org/abs/hep-th/0309228}{hep-th/0309228}.

\bibitem{KadanoffC71}
Kadanof\/f L.P., Ceva H., Determination of an operator algebra for
the two-dimensional Ising model, {\it Phys. Rev. B} {\bf 3} (1971),
3918--3939.

\bibitem{SchroerT78}
Schroer B., Truong T.T., The order/disorder quantum f\/ield
operators associated with the two-dimensional Ising model in the
continuum limit, {\it Nuclear Phys. B} {\bf 144} (1978), 80--122.

\bibitem{DoyonLectures05}
 Doyon B., Lectures on integrable quantum f\/ield theory,\\
 \url{http://www-thphys.physics.ox.ac.uk/user/BenjaminDoyon/lectures.pdf}.

\bibitem{vanHove86}
 van Hove L., Quantum f\/ield theory at positive temperature, {\it
Phys. Rep.} {\bf 137} (1986), 11--20.

\bibitem{Doyon03}
Doyon B., Two-point functions of scaling f\/ields in the Dirac
theory on the Poincar\'e disk, {\it Nuclear Phys. B} {\bf 675} (2003),
607--630,
\href{http://arxiv.org/abs/hep-th/0304190}{hep-th/0304190}.

\bibitem{McCoyWu}
McCoy B.M., Wu T.T., The two-dimensional Ising model,  Harvard University Press, Cambridge
(MA), 1973.

\bibitem{BabelonB92}
Babelon D., Bernard  D., From form factors to correlation
functions: the Ising model, {\it Phys. Lett. B} {\bf 288} (1992),
113--120.

\bibitem{LeclairLSS96}
Leclair A., Lesage F., Sachdev S., Saleur H., Finite temperature
correlations in the one-dimensional quantum Ising model, {\it
Nuclear Phys. B} {\bf 482} (1996), 579--612,
\href{http://arxiv.org/abs/cond-mat/9606104}{cond-mat/9606104}.

\bibitem{ItzyksonDrouffe}
Itzykson C., Drouf\/fe J.-M., Statistical f\/ield theory,
 Cambridge University Press, Cambridge, 1989.

\bibitem{Lisovyy05}
Lisovyy O., Tau functions for the Dirac operator on the cylinder,
{\it Comm. Math. Phys.} {\bf 255} (2005), 61--95,
\href{http://arxiv.org/abs/hep-th/0312277}{hep-th/0312277}.


\end{thebibliography}
\end{document}